\documentclass{elsart}
\usepackage{graphicx}

\journal{Nuclear Instruments and Methods A}

\begin{document}
\begin{frontmatter}

\title{DIRAC: A High Resolution Spectrometer for Pionium Detection} 

\author[p]{B.~Adeva\corauthref{cor1}},
%\footnote{\small{Corresponding author. 
%Departamento de  F\'{\i}sica de  Part\'{\i}culas, Universidade de Santiago de Compostela,
%E-15782 Santiago, Spain. Tel.: 34-981-563100,
%e-mail: adevab@usc.es}} 
\author[l]{L.~Afanasyev},
\author[d]{M.~Benayoun},
\author[q]{A.~Benelli},
\author[b]{Z.~Berka},
\author[o]{V.~Brekhovskikh},
\author[m]{G.~Caragheorgheopol},
\author[b]{T.~Cechak},
\author[j]{M.~Chiba},
\author[e]{E.~Cima},
\author[m]{S.~Constantinescu},
\author[a]{C.~Detraz},
\author[f]{D.~Dreossi},
\author[a]{D.~Drijard},
\author[l]{A.~Dudarev},
\author[s]{I.~Evangelou},
\author[a]{M.~Ferro-Luzzi},
\author[p,a]{M.V.~Gallas},
\author[b]{J.~Gerndt},
\author[f]{R.~Giacomich},
\author[e]{P.~Gianotti},
\author[e]{M.~Giardoni},
\author[q]{D.~Goldin},
\author[p]{F.~G\'omez},
\author[o]{A.~Gorin},
\author[l]{O.~Gortchakov},
\author[e]{C.~Guaraldo},
\author[a]{M.~Hansroul},
\author[e,m]{M.~Iliescu},
\author[l]{M.~Zhabitsky},
\author[l]{V.~Karpukhin},
\author[b]{J.~Kluson},
\author[g]{M.~Kobayashi},
\author[s]{P.~Kokkas},
\author[l]{V.~Komarov},
\author[l]{V.~Kruglov},
\author[l]{L.~Kruglova},
\author[l]{A.~Kulikov},
\author[l]{A.~Kuptsov},
\author[o]{V.~Kurochkin},
\author[l]{K.-I.~Kuroda},
\author[f]{A.Lamberto},
\author[a,e]{A.~Lanaro},
\author[o]{V.~Lapshin},
\author[c]{R.~Lednicky},
\author[d]{P.~Leruste},
\author[e]{P.~Levisandri},
\author[p]{A.~Lopez Aguera},
\author[e]{V.~Lucherini},
\author[i]{T.~Maki},
\author[s]{N.~Manthos},
\author[o]{I.~Manuilov},
\author[a]{L.~Montanet},
\author[d]{J.-L.~Narjoux},
\author[a,l]{L.~Nemenov},
\author[l]{M.~Nikitin},
\author[p]{T.~N\'u\~nez Pardo},
\author[h]{K.~Okada},
\author[l]{V.~Olchevskii},
\author[e]{D.~Orecchini},
\author[p]{A.~Pazos},
\author[m]{M.~Pentia},
\author[f]{A.~Penzo},
\author[a]{J.-M.~Perreau},
\author[e,m]{C.~Petrascu},
\author[p]{M.~Pl\'o},
\author[m]{T.~Ponta},
\author[m]{D.~Pop},
\author[f]{G.F.Rappazzo},
\author[o]{A.~Riazantsev},
\author[p]{J.M.~Rodriguez},
\author[p]{A.~Rodriguez Fernandez},
\author[p]{A.~Romero},
\author[o]{V.~Rykalin},
\author[p,q]{C.~Santamarina},
\author[p]{J.~Saborido},
\author[r]{J.~Schacher},
\author[q]{Ch.P.~Schuetz},
\author[o]{A.~Sidorov},
\author[c]{J.~Smolik},
\author[q]{M.~Steinacher},
\author[h]{F.~Takeutchi},
\author[l]{A.~Tarasov},
\author[q]{L.~Tauscher},
\author[p]{M.J.~Tobar},
\author[s]{F.~Triantis},
\author[n]{S.~Trusov},
\author[l]{V.~Utkin},
\author[p]{O.~V\'azquez Doce},
\author[p]{P.~V\'azquez},
\author[q]{S.~Vlachos},
\author[n]{V.~Yazkov},
\author[g]{Y.~Yoshimura},
\author[l]{P.~Zrelov}

\corauth[cor1]{Departamento de F\'{\i}sica de Part\'{\i}culas,
  Universidade de Santiago de Compostela, E-15782 Santiago, Spain.
  Tel.: 34-981-563100, e-mail: adevab@usc.es}

\address[a]{CERN, Geneva, Switzerland}
\address[b]{Czech Technical University, Prague, Czech Republic}
\address[c]{Institute of Physics ACSR, Prague, Czech Republic}
\address[s]{Ioannina University, Ioannina, Greece}
\address[d]{LPNHE des Universites Paris VI/VII, IN2P3-CNRS, France}
\address[e]{INFN - Laboratori Nazionali di Frascati, Frascati, Italy}
\address[f]{INFN-Trieste and Trieste University, Trieste, Italy}
\address[g]{KEK, Tsukuba, Japan}
\address[h]{Kyoto Sangyou University, Japan}
\address[i]{UOEH-Kyushu, Japan}
\address[j]{ Tokyo Metropolitan University, Japan}
\address[l]{JINR Dubna, Russia}
\address[m]{National Institute for Physics and Nuclear Engineering IFIN-HH Bucharest, Romania}
\address[n]{Skobeltsin Institute for Nuclear Physics of Moscow State University Moscow, Russia}
\address[o]{IHEP Protvino, Russia}
\address[p]{Santiago de Compostela University, Spain}
\address[q]{Basel University, Switzerland}
\address[r]{Bern University, Switzerland}

\begin{abstract}

The DIRAC spectrometer has been commissioned at CERN with the
aim of detecting $\pi^+ \pi^-$ atoms produced by a 
24~GeV/$c$ high intensity proton beam 
in thin foil targets. A challenging apparatus is required to cope with
the high interaction rates involved, the triggering of
pion pairs with very low relative momentum, and the measurement
of the latter with resolution around 0.6~MeV/$c$.  
The general characteristics of the apparatus are
explained and each part is described in some detail. The main
features of the trigger system, data-acquisition, monitoring and 
setup performances are also given.

\end{abstract}
\vspace{.5cm}


\end{frontmatter}

\newpage

\section{Introduction}

The DIRAC experiment aims to measure the ground state lifetime of
$\pi^+ \pi^-$ atoms with 10\% precision, using the 24~GeV/$c$ proton
beam of the CERN Proton Synchrotron.  The atom lifetime is a
consequence of the strong interaction at low energy and it is
determined by the charge exchange amplitude $\pi^+ \pi^- \rightarrow
\pi^0 \pi^0 $ very close to threshold.  The probability of this
process is proportional to the square of the difference of S-wave $\pi
\pi$ scattering lengths with isotopic spin 0 and 2, ${|a_0 - a_2|}^2$.
The relation between the lifetime and $|a_0 - a_2|$ is
model-independent \cite{lyub}. The pion scattering lengths have been
calculated in the framework of chiral perturbation theory with a
precision of a few percent: $ a_0 = 0.220 \pm 0.005$ and $a_2 =
-0.0444 \pm 0.0010 $.  Using these values one can predict the pionium
lifetime \cite{gasser}: $(2.9 \pm 0.1) \times 10^{-15} s $. In order
to determine $|a_0-a_2|$ down to 5\%, the lifetime has to be measured
within 10\% accuracy. Such a measurement would provide a crucial test
for the understanding of chiral symmetry breaking in QCD.

Pionium atoms ($A_{2\pi}$) are produced in proton-nucleus
interactions. After production these relativistic atoms may either
decay into $\pi^0 \pi^0$ or get excited to higher quantum numbers, or
break up (be ionised) in the target material where they are produced.
In the case of break-up, characteristic pion pairs (``atomic'' pairs)
emerge. These pairs have a low relative momentum in their centre of
mass system ($ Q < 3$~MeV/$c$), very small opening angle ($\theta <3
$~mrad) and nearly identical energies in the laboratory system.  A
high resolution magnetic spectrometer is then required \cite{dirac} to
split up the pairs and measure their relative momentum with sufficient
precision (0.6~MeV/$c$) to detect the pionium signal superimposed on
the substantial background of ``free'' $\pi^+ \pi^-$ pairs produced in
inclusive proton-nucleus interactions.  A previous experiment, using
internal proton beam, has reported observation of pionium atoms
\cite{sherp}.

The total number of produced $\pi^+ \pi^-$ atoms is related by an
exact expression to the number of free pion pairs with low relative
momenta. For a given target material and thickness the ratio of
observed atomic pairs to the total number of produced atoms, i.e. the
atom breakup probability, depends on the lifetime in a unique way
\cite{neme}.

\section{General layout of the experimental setup}

The DIRAC experimental setup \cite{kup1,kup2} is located at the T8
proton beam line of 24~GeV/$c$ momentum in the East Hall of the PS
accelerator at CERN.  The isometric view of the setup is shown in
Fig.\ref{fig:c3iso}.  The DIRAC apparatus is designed to detect
charged pion pairs with high resolution over the pair relative
momentum. It became operational at the end of 1998 and has been
collecting data since the middle of 1999.

\begin{figure}[htb]
\begin{center}
\caption{ Isometric view of the DIRAC setup.
  The radiation shielding boundaries are shown on the floor (each
  division marked on the boundary corresponds to 1~meter).}
\label{fig:c3iso}
\end{center}
\end{figure}

\begin{figure}[htb]
\begin{center}
\caption{ Side view of the DIRAC setup. The secondary particle channel
  is inclined by $5.7^\circ$ with respect to the primary proton beam.}
\label{fig:c2side}
\end{center}
\end{figure}

The setup consists of the proton beam line, target station, secondary
particle vacuum channel, spectrometer magnet and detectors placed
upstream and downstream the analysing magnet.  Free and atomic $\pi^+
\pi^-$ pairs produced in the target enter the secondary particle
channel which is tilted upwards by 5.7$^\circ$ with respect to the
proton beam (Fig.\ref{fig:c2side}).  At the end of the secondary
particle channel the spectrometer magnet is installed, also tilted by
5.7$^\circ$ together with all the downstream detectors.
 
\begin{figure}[htb]
\begin{center}
  \includegraphics*[width=\textwidth]{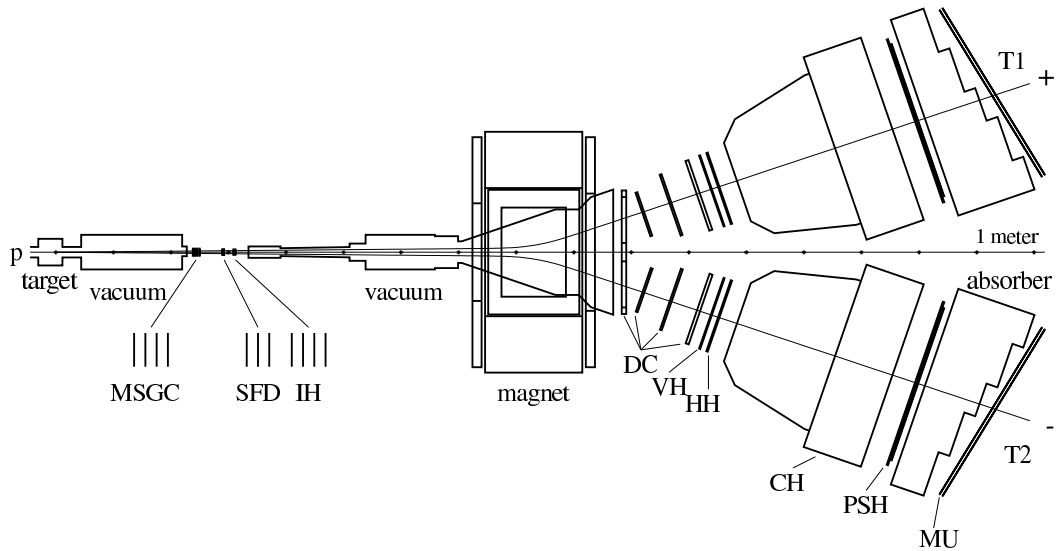}
\caption{\label{fig:c0top} 
  Schematic top view of the DIRAC spectrometer. Moving from the target
  station towards the magnet there are 4 planes of microstrip gas
  chambers (GEM/MSGC), 3 planes of scintillating fibre detectors (SFD)
  and 4 planes of ionisation hodoscope (IH). Downstream the dipole
  magnet, on each arm of the spectrometer, there are 4 stations of
  drift chambers (DC), vertical and horizontal scintillation
  hodoscopes (VH, HH), gas Cherenkov counter (CH), preshower detector
  (PSH) and, behind the iron absorber, muon detector (MU).}
\end{center}
\end{figure}

The top view of the setup is shown in Fig.\ref{fig:c0top}.  The
upstream section of the secondary particle channel between the target
station and the spectrometer magnet is instrumented with the following
detectors: microstrip gas chambers (GEM/MSGC), scintillating fibre
detector (SFD) and scintillation ionisation hodoscope (IH).

Downstream the spectrometer magnet the setup splits into two identical
arms for detection and identification of positive and negative charged
particles.  The angle between each arm and the spectrometer symmetry
axis is 19$^\circ$.  Along each arm the following detectors are
located: drift chamber system (DC), vertical scintillation hodoscope
(VH), horizontal scintillation hodoscope (HH), gas Cherenkov counter
(CH), preshower detector (PSH) and muon detector (MU).

\section{Beam lines, spectrometer magnet and radiation 
  shielding}

\subsection{Proton beam and target station}

To extract protons from the PS to the T8 beam line a slow ejection
mode is used. The beam is extracted in spills of $\approx400$--500~ms
duration.  During data taking, between 1 to 5 cycles per PS
super-cycle of $14.4\div19.2$~s duration are delivered to DIRAC.  The
proton beam intensity was set to $(0.6-1.0)\cdot10^{11}$ protons per
spill, depending on the target used.

The PS proton beam line includes two bending magnets deflecting the
beam at an angle of 76~mrad towards the final straight T8 section,
corrector magnets performing horizontal and vertical steering and
quadrupole magnets which focus the beam on the experiment target.  The
dimension of the beam spot at the target location are $x=1.6$~mm,
$y=3.2$~mm at $2\sigma$ level. The divergence of the beam is about
1~mrad.  The nominal momentum of the extracted beam is 24~GeV/$c$,
with instantaneous momentum spread close to 0.08\% at 2$\sigma$.  The
design of the proton beam line optics has been optimised using the
TRANSPORT simulation code \cite{TRANSPORT} by the CERN PS Division.

Downstream the target the proton beam travels in a vacuum channel
below the spectrometer magnet and detectors and finally is absorbed by
a beam dump.

To measure the beam intensity and to tune the beam position on the
target, the beam line is equipped at several locations with secondary
emission chambers and luminescent screens with TV cameras.  One
additional beam position detector (centroid) \cite{BOSS02} is
installed close to the target station.

The target station houses a remote controlled device with 12 holders
for the targets, including an empty holder and a luminescence screen.
During data taking, targets made of Pt (28~$\mu$m thick), Ni
(94~$\mu$m and 98~$\mu$m thick) and Ti (250~$\mu$m thick) were used.

The DIRAC experiment is sensitive to particles outside the beam core
(halo), because the target is very thin (nuclear target efficiency is
$< {10}^{-3}$), and the upstream detectors are placed very close
(18--26~cm in the vertical direction) to the primary proton beam.  The
halo is originated from scattering of primary protons on the splitter
blades, and a special optics has been designed to decrease the
background halo to a negligible level.  The ratio of detector counting
rates with the target in place to those with an empty holder was
measured to be $\sim$25.

\subsection{Secondary particle channel and spectrometer magnet}

The secondary particle channel \cite{kup1} \cite{kup2} is placed at an
angle of 5.7$^\circ$ relative to the proton beam and consists of two
vacuum volumes, as shown in Fig.~\ref{fig:c0top}.  The first one is a
2~m long, 611~mm diameter tube, located immediately downstream the
target station, common to both the proton beam line and the secondary
particle channel.  Secondary particles exit this tube through a 200~mm
diameter window, made of 250~$\mu$m thick mylar film.  The second
volume, located at $\sim $3.5~m from the target, consists of a
cylindrical vacuum section, containing a collimator, attached to a
2.7~m long flat vacuum chamber placed between the spectrometer magnet
poles.  A 1.5~m long air gap between the two vacuum sections allows
insertion of the upstream detectors.  The angular aperture of the
secondary particle channel is determined by the collimator and is
equal to $\pm 1^\circ$ in horizontal and vertical directions resulting
in a solid angle acceptance of $1.2\cdot10^{-3}$~sr.  The flat chamber
is ended with a 0.68~mm thick Al outlet window of $2.0\times0.4$~m$^2$
dimensions ($W \times H$).

The spectrometer dipole magnet (magnetic field B=1.65~T, field
integral BL${}=2.2$~T$\cdot$m) has an aperture of
$1.55\times0.50$~m$^2$ ($W \times H$). To reduce the stray field, two
magnetic screens are fixed near its entrance and exit.
 
Table~\ref{tabchan} summarises the material thicknesses (in units of
radiation length $\times 10^{-4}$) encountered by secondary particles
before they reach the DC system where their momenta are measured.

\begin{table}[htbp]
\caption{ Material contributions along the
secondary particle channel in units of radiation length $\times 10^{-4}$.}
\label{tabchan}
\begin{center}
\begin{tabular}{|l|r|}
\hline
   Ni-target              &   33.5 \\ \hline
   Mylar window           &    8.7 \\  
   4 planes GEM/MSGC      &  224.1 \\ 
   3 planes SFD           &  260.0 \\ 
   4 planes IH            &  153.1 \\ 
   air gap                &   34.7 \\ 
   Mylar window           &    8.7 \\ 
   Al-window              &   76.4 \\ 
   Total                  &  765.7 \\ \hline
\end{tabular}
\end{center}
\end{table}

\subsection{Beam dump and radiation shielding}

\begin{figure}[htb]
\begin{center}
  \includegraphics*[width=\textwidth,bb=40 117 525 270]{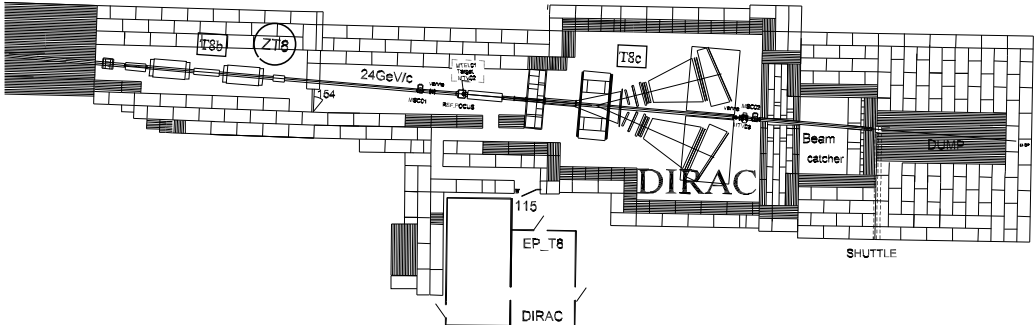}
\caption{\label{fig:b1ehl}
  The DIRAC setup on the T8 PS extraction line and the radiation
  shielding.}
\end{center}
\end{figure}

Neutron and gamma fluxes may cause a serious problem to sensitive
elements of the setup. To estimate their effect a simulation of the
background radiation flux in the full experimental apparatus has been
performed \cite{KURO96}. The results were used to optimise the design
of the radiation shielding which is shown in Fig.~\ref{fig:b1ehl}.

At the end of the T8 beam line primary protons are absorbed by an iron
beam dump. To decrease the background gamma and neutron fluxes from
the beam dump towards the detectors, a dedicated radiation shielding
has been adopted.  It includes installation of a graphite core into
the beam dump area, a concrete wall near the beam dump and, at 3~m
distance, another iron-concrete-iron wall, both with holes to allow
passage of the proton beam pipe.

The downstream detectors are shielded in addition from background
secondary particles produced on the primary proton pipe and
surrounding elements. For this purpose a 1~m thick iron wall is
installed between the upstream detector region and the spectrometer
magnet.  In addition, collimators are inserted both in the primary
proton beam pipe and in the secondary particle channel.  The presence
of the collimator in the proton beam line determines a reduction of
the background rate by a factor of 2.

A radiation shielding encloses the whole DIRAC experimental apparatus
to protect the surrounding East Hall area from irradiation.  It has
been designed according to the maximum flux of $2.7 \times {10}^{10}$
incident protons per second, in respect of the CERN safety
regulations.  Being the apparatus located in a fully enclosed area, it
has become necessary to provide the experimental area with cooling and
ventilation equipment to prevent overheating of detectors and
electronics.

\section{Large Detectors upstream the magnet}

Two tracking devices have been installed in the secondary particle
channel before the magnet and the collimator : the GEM/MSGC and the
SFD. They are used to improve the resolution on the measurement of the
longitudinal and transverse components of the relative momentum of
pion pairs as determined by the drift chambers tracking system and the
nominal position of the beam at the target center.

At the same time these detectors allow to select particle pairs
originated by primary interactions at the target from the background
of secondary interactions and particle decays. The ensemble of
MSGC/GEM+SFD constitutes a tracking system with 7 detector planes with
2 stereo angles which provides adequate space resolution to reach the
limit of multiple scattering in the target material.

To increase the detection capability on close-lying tracks, an
Ionisation Hodoscope (IH) is installed downstream the SFD, with the
purpose of detecting pion pairs with a too small opening angle to be
resolved by the tracking detectors. This is achieved by a detailed
pulse-hight analysis of the double ionisation produced by the particle
pairs in 4 layers of scintillation counters.

A general picture of the above-mentioned detectors, as they are
installed between the first vacuum chamber and the secondary particle
channel, can be seen in Figure \ref{up_photo}.

\begin{figure}[htb]
\begin{center}
\caption{Photography of the three detectors installed upstream the magnet,
  between the first vacuum chamber (right-hand side) and the secondary
  particle channel. From right to left, the GEM/MSGC, SFD and IH
  detectors can be found. The primary proton beam line can be
  appreciated at the bottom.}
\label{up_photo}
\end{center} 
\end{figure}

\section{The GEM/MSGC detector}

This detector performs particle tracking at a distance of 2.4~m from
the interaction point. It is a proportional gas detector, based on the
principle of the Gas Electron Amplifier (GEM) \cite{gemref}
\cite{teresa}, complemented with a second amplification and readout
stage provided by Micro Strip Gas Chambers (MSGC) \cite{oed}
\cite{nimus}.  A more complete description of the detector and its
performance is being prepared in a separate publication
\cite{nimprep}.
 
It measures particle coordinates in 4 planes along the direction of
the incoming particle: X,Y,U,V, with orientations 0, 90, 5, 85
degrees, respectively, where the 0~degrees are defined by microstrips
running vertically (X-coordinate). The stereo angles allow resolution
of ghost combinations for two or more particles.  With a single-hit
space resolution close to 54~$\mu$m, this detector provides a precise
measurement of the pion pair angular opening, ultimately limited by
multiple scattering in the thin target.

\subsection{Detector concept}

\begin{figure}[htb]
\begin{center}
  \includegraphics*[width=0.93\textwidth]{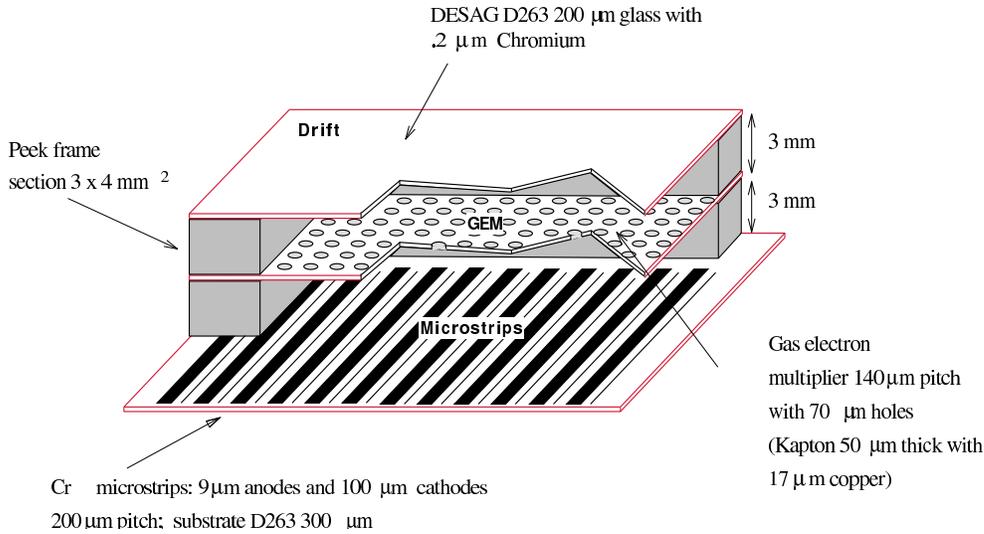}
\caption{\label{gem1}  Layout of the GEM/MSGC detector.} 
\end{center} 
\end{figure}

Each chamber has active area 10.24$\times$10.24~cm$^2$, and consists
of a drift electrode, a GEM foil and a MSGC sensor. The GEM plane is
evenly spaced from the other two with a uniform gap of 3~mm, as
indicated in Fig.~\ref{gem1}. The drift electrode is made of a
Chromium-coated thin glass (200~$\mu$m). The GEM is a 50~$\mu$m thick
kapton foil copper-cladded on both sides with a 4~$\mu$m thick Cu
layer. The etching pattern is characterised by 50~$\mu$m wide holes,
140~$\mu$m apart \cite{gandi}. Application of a potential difference
of 400~V between the two metal layers ($V_1 = -1600$~V, $V_2 =
-2000$~V) produces electron amplification by a factor of 30
\cite{teresa}.
 
The MSGC sensor consists of 200~$\mu$m pitch alternating Chromium
strips, with 9~$\mu$m and 100~$\mu$m anode and cathode width,
respectively. They are implanted on a bare DESAG D263 substrate.
Applied voltages are: $-410$~V on cathodes, and $-3000$~V on the drift
electrode, whereas anode strips are set to ground. The gas employed is
a mixture of Ar-DME (60/40).  Under these conservative conditions an
overall detector gain of approximately 3000 is achieved.

\subsection{Readout electronics}

It is composed of three parts: the analog boards, the control boards
and the VME acquisition modules \cite{pablo}.

The analog boards contain the front-end electronics, and also serve as
a mechanical support for the detector.  They host 16 Analog Pipeline
Chips (APC) \cite{horis}, bonded on a multi-layer hybrid to the
detector fan-out strips.  Each APC has 64 inputs, with a charge
sensitive preamplifier followed by analog pipeline of 32 capacitors,
which is run at 10~MHz. In order to accelerate the serial readout,
only 32 inputs are connected to the same chip. The analog pulse is
formed by charge subtraction from consecutive capacitors, at positions
given by the delayed trigger signal. This robust procedure, which
allows easy integration of a large number of channels, limits however
the time resolution of the readout to a window of approximately
200~ns.

The control boards store the digital sequence needed for APC control
and perform fast digital conversion (AD9048) of the 16 series of
analog signals. These two functions are controlled by FPGA (Fast
Programmable Gate Array) incorporated onto the board.

The VME modules handle the trigger signal and perform real time
pedestal subtraction and zero suppression. One module is divided into
4 segments, each controlled by an independent FPGA. A Fast Clear
function is implemented to be used in connection with the highest
level of trigger (see trigger chapter), that stops the readout of
current event and enables new triggers.

\begin{figure}[htbp]
\begin{center}
  \includegraphics*[height=7cm,bb=25 20 515 630]{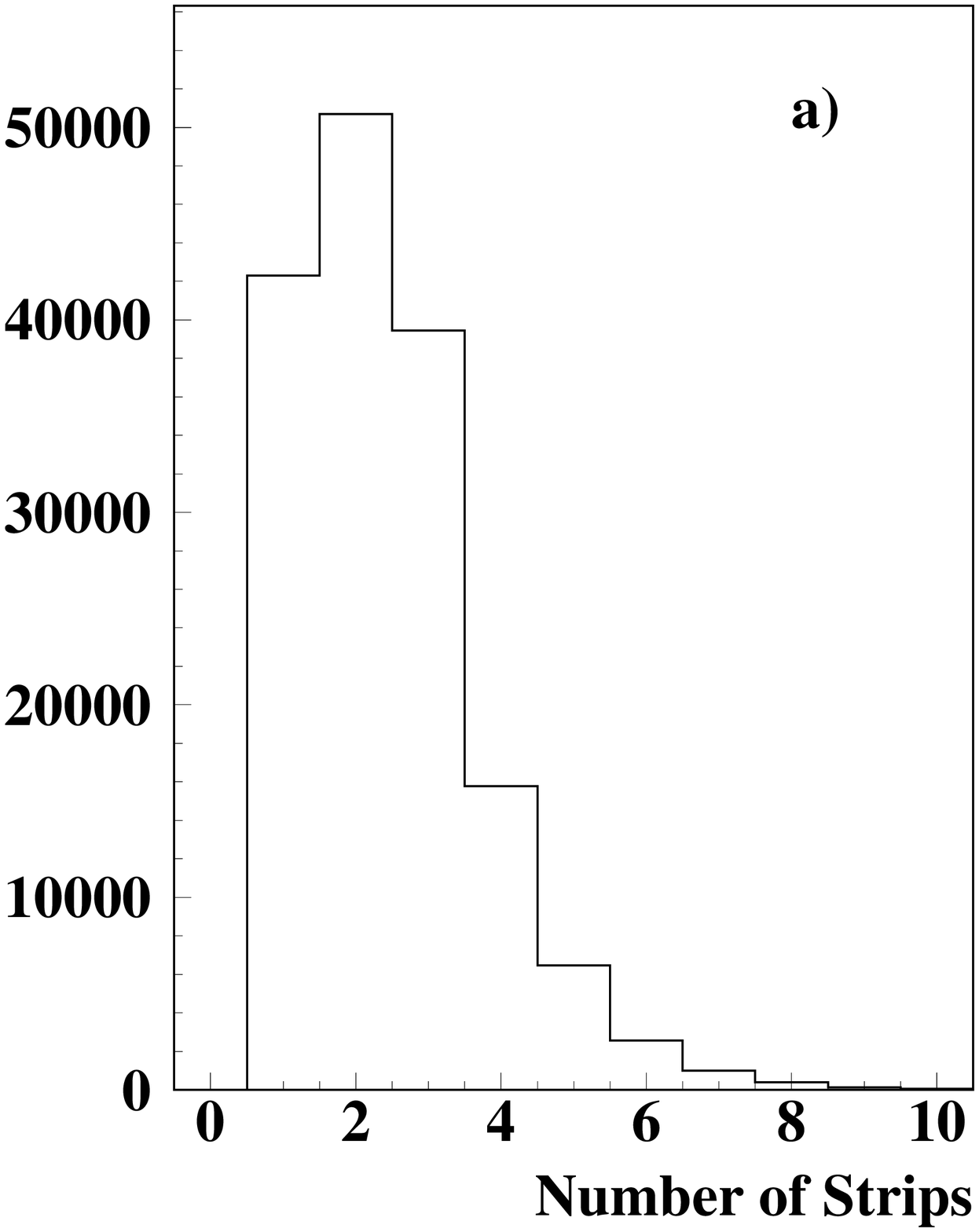}~~~
  \includegraphics*[height=7cm,bb=25 20 515 630]{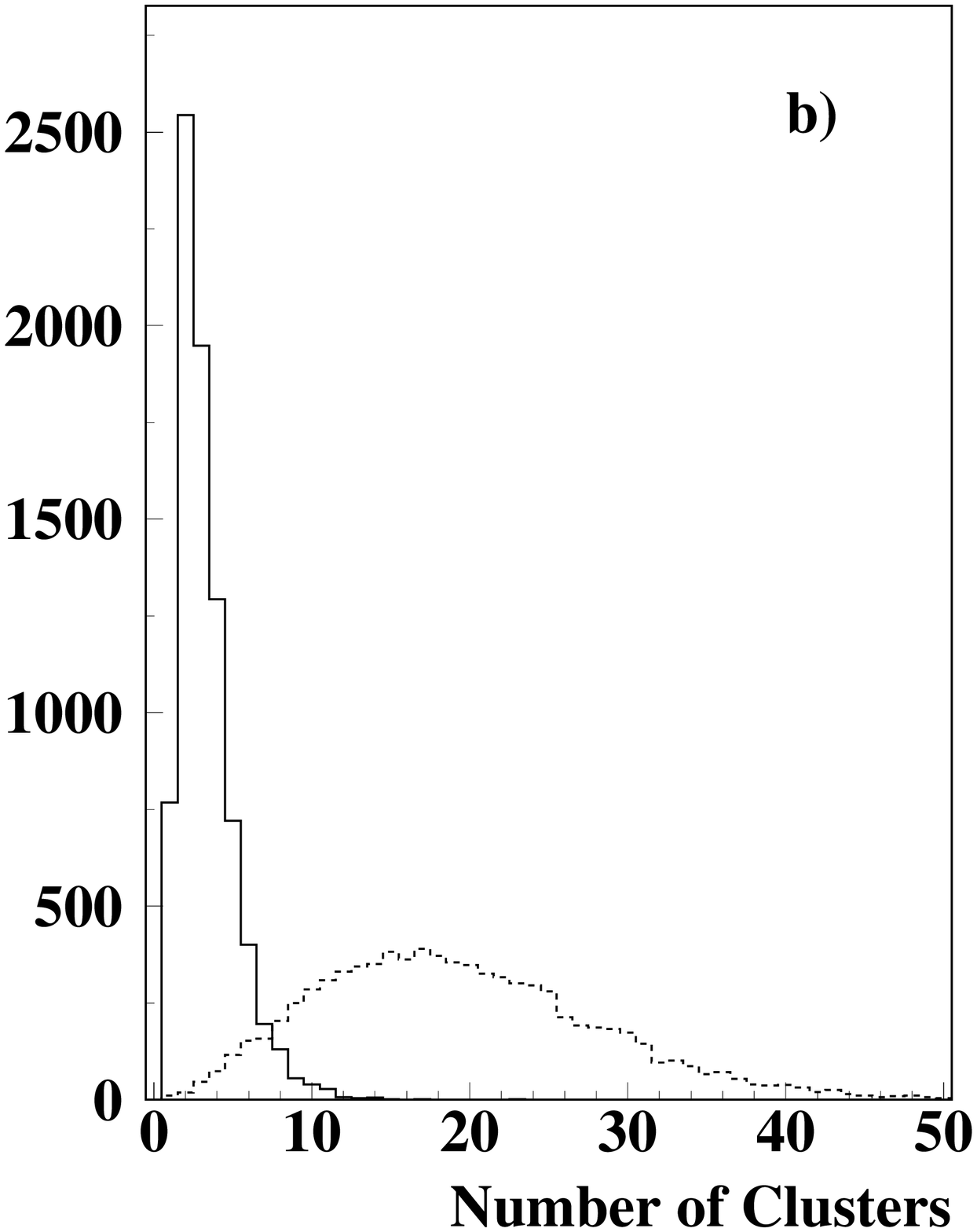}
\end{center} 
\caption{a) GEM/MSGC strip multiplicity per hit-cluster
  b) cluster multiplicity per event in one plane (X). The dotted line
  shows all clusters registered by the data acquisition, whereas the
  continuous line shows only those having a time tag, established when
  the cluster has a corresponding hit in the SFD aligned with the
  interaction point.}
\label{gem2}
\end{figure} 

\subsection{Performance}

\begin{figure}[htbp]
\begin{center}
  \includegraphics*[width=\textwidth,height=\textwidth,bb=25 25 400
  405]{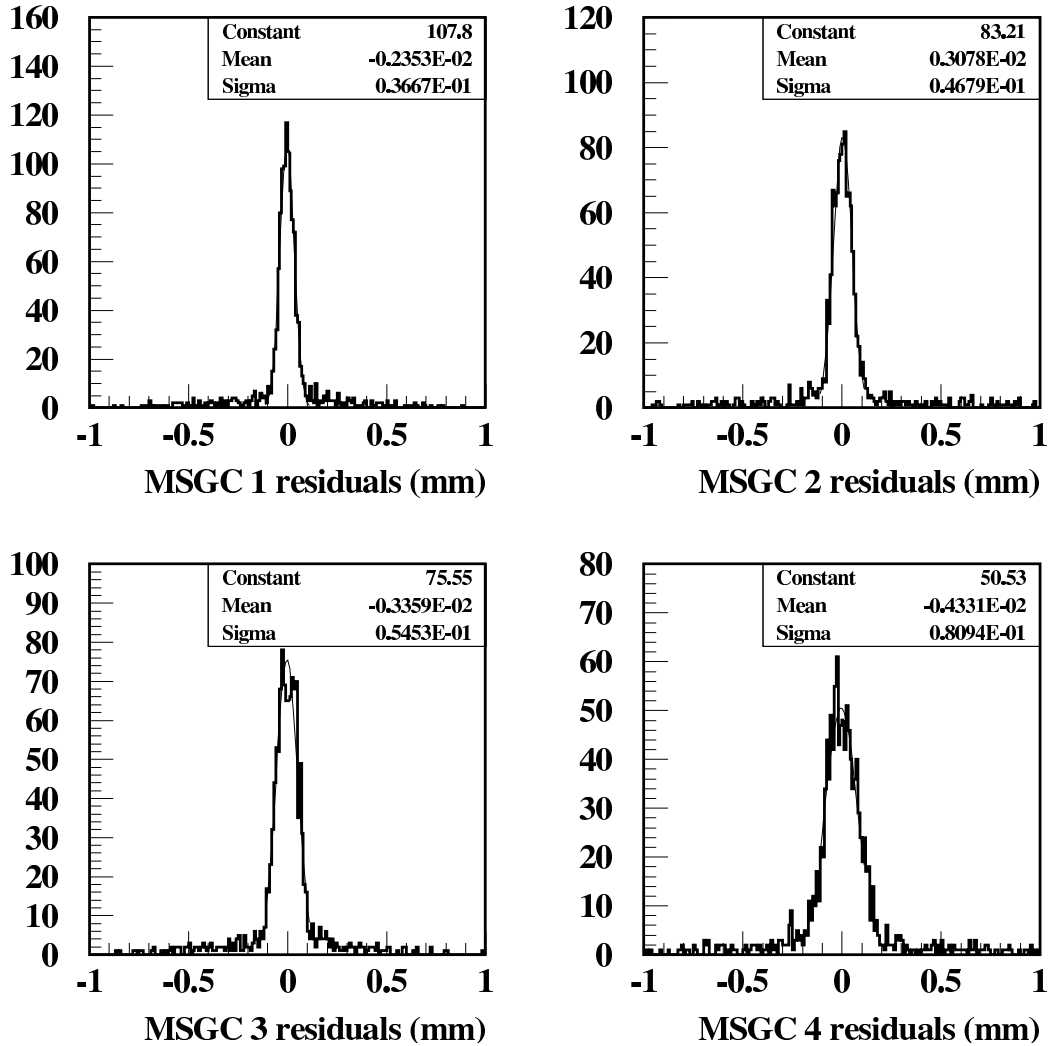}
\caption{\label{gem3}  Space resolution of the GEM/MSGC detector measured
  in a dedicated beam-test with 4 planes parallel to each other.}
\end{center} 
\end{figure}

Neighbouring hit strips are pattern-recognised as clusters.  The hit
multiplicity in them (a) and the cluster multiplicity per plane (b)
are shown in Fig.~\ref{gem2}.  Single hit resolution was determined
during the commissioning run in April 2000 by setting all planes
parallel to each other, and 54~$\mu$m is a typical value, as shown in
Fig.~\ref{gem3}.  Two detector planes were installed in 1999 and the
full set in 2000.  The number of detector dead channels is around 1\%
and no significant deterioration due to radiation has been observed
since then.  The average efficiency for a standard detector is 93\%.
When at least 4 signals among 7 detectors are required to make a
track, this is sufficient to provide 99\% overall tracking efficiency
upstream the magnet.

\section{The Scintillating Fibre Detector}

The Scintillating Fibre Detector (SFD), together with the GEM/MSGC
detector, enables particle tracking to be performed upstream the
magnet, with the required space and time resolution.  In addition, it
provides topological trigger capabilities \cite{SFDtrig} for rejection
of pairs with relative distances larger than 9~mm at the detector
location (this feature was used at the early stage of the experiment,
see a footnote in section ``Trigger system'').

\subsection{Detector concept}

\begin{figure}[htb]
\begin{center}
  \includegraphics*[width=0.85\textwidth]{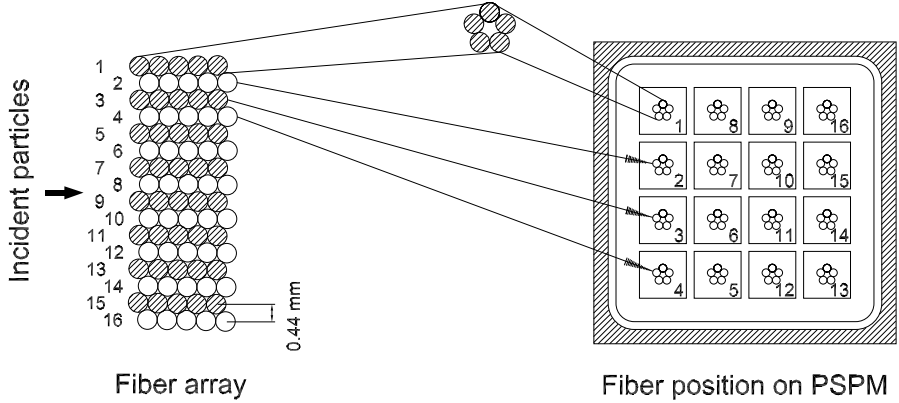}
\end{center}
\caption{The SFD principal structure. A 16-channel fragment is shown.}
\label{sfdstructure}
\end{figure}

The SFD consists of three fibre planes to measure the X-, Y- and
U-coordinates of incident particles\footnote{The SFD operational in
  DIRAC until the end of 2001 run consisted of X and Y planes only.
  The U-plane described here was installed in early 2002.}.  The SFD
covers a 105x105~mm$^2$ area, each scintillating fibre (SciFi) array
consists of several layers of KURARAY fibres. Five (for planes X and
Y) or three (for plane U) fibres forming one sensitive column are
mapped onto one channel of position-sensitive photomultiplier (PSPM).
A layout illustrating the main characteristics of the SFD (for a plane
with 5 fibres per column) and PSPM mapping is shown in
Fig.~\ref{sfdstructure}.

The fibre columns pitch is 0.44~mm, thus allowing some overlap to
optimise efficiency.  Each SciFi is connected by optical epoxy to a
clear fibre light guide of $\sim$300~mm length.  The edges of clear
fibres, assembled in a bundle, are carefully polished and glued into
the holes of a square black plate which fixes the fibre positions on
the PSPM photocathode; no optical grease is used. The far end of the
SciFi array is tightly connected to a mirror made of aluminised mylar.
With this arrangement the light attenuation along the SciFi was found
to be negligible.  The details of the fibre arrays are listed in
Table~\ref{tab1}.

\begin{table}[htbp]
\begin{center}
  \caption{SFD planes specification.}\label{tab1}
\begin{tabular}{|c|c|c|c|c|c|c|c|}
\hline
   & Fibre      &  Fibre   & Length & Fibres/ & $X_0$ & Num. of & Num. of \\
   & type       &$\Phi$ (mm)&  mm   & column  &  \%    & channels  & PSPM \\ \hline
X & SCSF38      & 0.50     & 130     & 5       & 0.8  & 240       & 15 \\
Y & SCSF38      & 0.50     & 130     & 5       & 0.8  & 240       & 15 \\
U & SCSF78M     & 0.57     & 150,    & 3       & 1.0  & 320       & 20 \\ 
  &             &          & 130, 70 &         &      &           & \\ \hline
\end{tabular}
\end{center}
\end{table}

The SFD individual planes are installed close to each other.
Direction of fibres is orthogonal in planes X and Y while plane U is
rotated by 45 degrees.  The U-plane has 5 sections of different fibre
lengths to cover roughly the same area like the X- and Y-planes.
 
A 16-channel metal dynode position-sensitive photomultiplier tube,
Hamamatsu H6568, has been selected as photosensor.  This
photomultiplier is characterised by good timing properties (rise time
$\sim$0.7~ns), low noise (1--2 pulses/s at a nominal detection
threshold) and perfect single photoelectron spectrum.  It has been
modified to monitor the amplitude from the last dynode for calibration
purposes.  The level of optical cross talk among the PSPM channels was
found to be $\sim$1\% (with a 1.6~mm diameter light spot on the
photocathode).  The measured linear range of this tube (for the linear
bleeder) extends up to 15 photoelectrons at 950~V.
 
The detector concept with almost independent channels was developed in
the context of the RD-17 project \cite{RD17}. The results from the SFD
study with a test beam of low intensity, reported in
\cite{RD17results}, are the following:

\begin{itemize}
\item[---] light output 6--10 photoelectrons,
\item[---]  average detection efficiency 98.4\%,
\item[---]  r.m.s. of the detection efficiency 1\%,
\item[---]  average hit multiplicity 1.1,
\item[---]  spatial resolution ($\sigma$) 127~$\mu$m,
\item[---]  time resolution ($\sigma$) 0.65~ns.
\end{itemize}

\begin{figure}[htb]
\begin{center}
  \includegraphics*[width=0.7\textwidth]{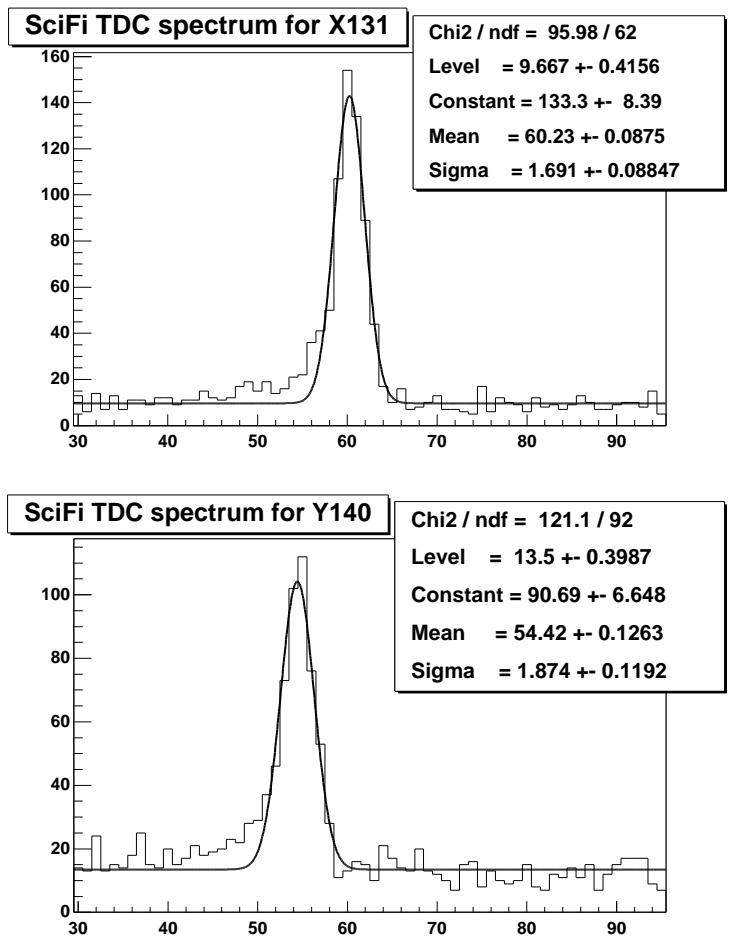}
\end{center}
\caption{SFD raw time spectra for $e^+e^-$ trigger data. Here and in
  Fig.~\ref{sfdpi} the horizontal scale is in TDC channels, the
  channel width is 0.5~ns.}
\label{sfdel}
\end{figure}

\subsection{Readout electronics}

A dedicated electronic circuit (PSC) has been custom developed to
provide signal discrimination with dynamic rejection of cross-talk in
adjacent channels using the peak-sensing technique.  A detailed
description of a 32-channel PSC module is given in \cite{peaksens}.
Discrimination of a channel is given by the condition $ 2 A_i -
A_{i-1} - A_{i+1} > A_{thr}$, where $A_{i}$ are channel signal
amplitudes and $A_{thr}$ defines the threshold value.  The PSC
algorithm provides efficient detection of double tracks from time
correlated particle pairs (up to $\sim$5~ns time difference) when the
relative distance between the two tracks is larger than the fibre
column pitch.  However, when adjacent fibre columns are crossed by two
particles simultaneously, then the PSC algorithm leads to a
suppression (by 20-40\%) of the detected yield of double track events.
In such cases a signal is detected for only one of two hit columns.
The detection efficiency depends on the detector light output in a
specific channel.  For particle pairs with relative time differences
greater than 10~ns (accidental pairs) the PSC behaves as an ordinary
leading edge discriminator.

\begin{figure}[htb]
\begin{center}
  \includegraphics*[width=0.7\textwidth]{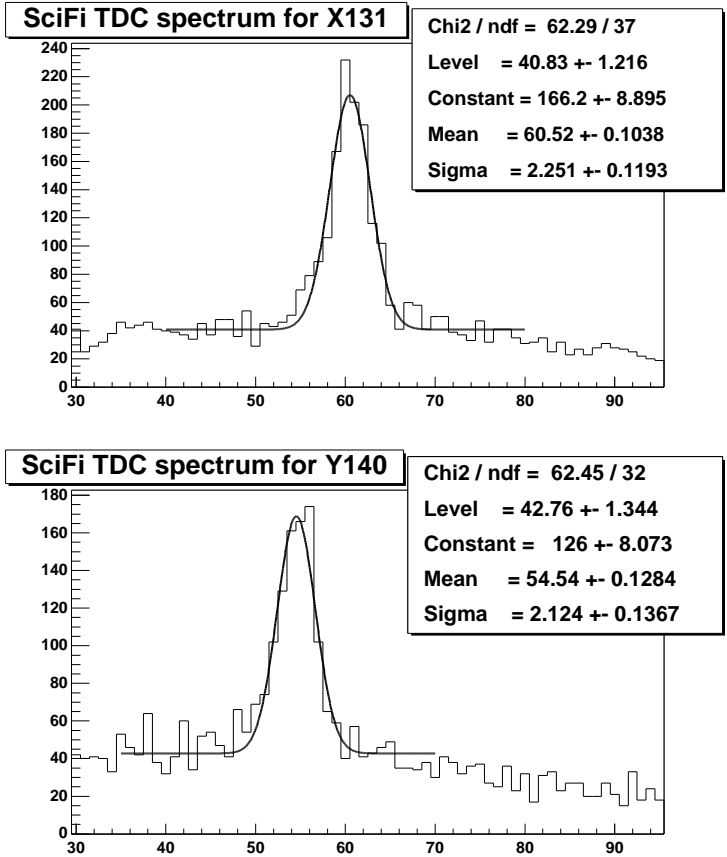}
\end{center}
\caption{SFD raw time spectra for $\pi^+ \pi^-$ trigger data.} 
\label{sfdpi}
\end{figure}

The front-end electronics is located inside the beam area close to the
detector. The PSC output ECL signals are sent to LeCroy 3377 multi-hit
TDC.  This solution does not require additional delay lines to adjust
the timing of the SFD with respect to the trigger timing.

\subsection{Performance in the experiment}

Due to the high flux of particles at the position of the detector
close to the target, and to the presence of a non-negligible amount of
inclined tracks associated with secondary interactions in the channel,
the SFD performances in the experiment slightly differ from the above
mentioned. The detection efficiency is still high (around 98\%), but
the average hit multiplicity is near 5 in the 50~ns time window of TDC
(at a nominal beam intensity of $10^{11}$ protons per spill impinging
on a 94~$\mu$m Ni target).  The raw time spectra, obtained from
$e^+e^-$ and $\pi^+ \pi^-$ events, are shown in Fig.~\ref{sfdel} and
Fig.~\ref{sfdpi}, respectively, for two arbitrary SFD channels.  The
width of the distributions is dominated by the time jitter of the
trigger signal. After off-line deconvolution of the trigger time
jitter the resolution of the SFD is found to be $\sigma$=0.8~ns.

\section{The Ionisation Hodoscope}

Charged pions originated from pionium breakup cross the upstream
detectors at rather small relative distances.  When the distance is
less than the double track resolution of the upstream tracking
devices, then only one hit is detected, thus making the event
reconstruction ambiguous. That is why another technique based on a
measurement of the ionisation loss is used as well.

A dedicated Ionisation Hodoscope (IH)~\cite{diracnote0902} has been
built to separate double ionisation signals produced by close pion
pairs incident on the same scintillating slab, from single ionisation
signals produced by one particle.  In this way, the uncertainties
resulting from the inefficiency in detecting two tracks with relative
distance approaching zero can be significantly reduced.

The Ionisation Hodoscope described here was installed in 2001 to
replace a previous version of a similar detector type
\cite{mfl_notes}, consisting of only two planes with 16 slabs of 2~mm
thickness, oriented in the vertical direction.

\begin{figure}[htb]
\begin{center}
  \includegraphics*[height=5cm,bb=190 70 405 370]{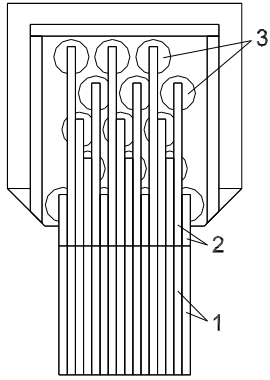}
\end{center}
\caption{Design of the IH scintillation plane. 1~--~scintillators,
  2~--~light-guides, 3~--~PM photocathodes.}
\label{ihdesign}
\end{figure}

\begin{figure}[htb]
\begin{center}
  \includegraphics*[height=5cm,bb=130 50 475 385]{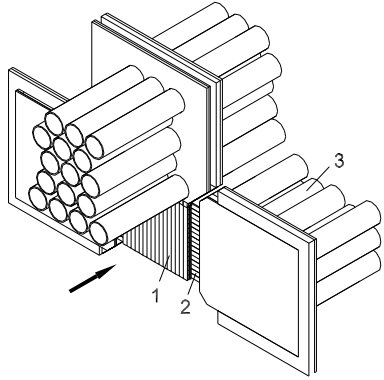}
\end{center}
\caption{Isometric view of the Ionisation Hodoscope.  1~--~scintillators, 
  2~--~light-guides, 3~--~photomultipliers with shielding.}
\label{planedesign}
\end{figure}

The present IH detector is a scintillation hodoscope consisting of
4~planes of $11\times 11\mbox{ cm}^2$ sensitive area placed normally
to the axis of the setup (Fig.~\ref{ihdesign}), 3~m downstream the
target.  Two planes have vertically oriented slabs (planes X-A and
X-B) whereas the other two have horizontal slabs (planes Y-A and Y-B).
They are arranged in the following sequence, moving along the beam
direction: X-A, Y-A, X-B, Y-B.  This ordering has been chosen to
minimise possible cross-correlations between signals in the planes
(e.g. due to $\delta$-electrons).  Each plane is assembled from 16
plastic scintillating slabs made of fast scintillator (BC-408).
Planes with the same slab orientation are shifted by a half-slab-width
with respect to each other.  The slabs are 11~cm long, $7$~mm wide and
1~mm thick.  They are connected to the PM photocathodes via 2~mm thick
and 7~mm wide lucite light guides (fig.~\ref{planedesign}).

The front and rear surfaces of a slab are covered by a millipore
film~\cite{millipore} for efficient light collection.  At the lateral
surface of the slab, light is reflected by a thin (30~$\mu$m)
aluminised black mylar film, which is used instead of millipore film
in order to minimise the gaps between adjacent slabs.  A typical gap
between two adjacent slabs in this configuration is less than
70~$\mu$m wide.

\begin{figure}[htb]
\begin{center}
  \includegraphics*[width=10cm]{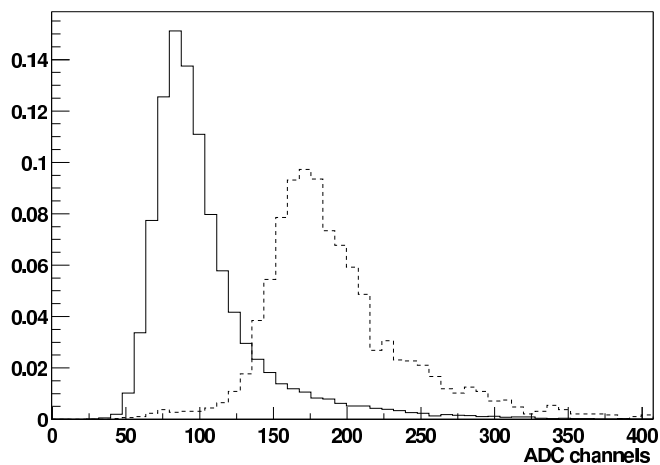}
\end{center}
\caption{ Typical ADC spectra for single (solid line) and
  double (dashed line) ionisation loss from particles crossing one IH
  scintillating slab.}
\label{adcspectra}
\end{figure}

\begin{figure}[htb]
\begin{center}
  \includegraphics*[width=10cm]{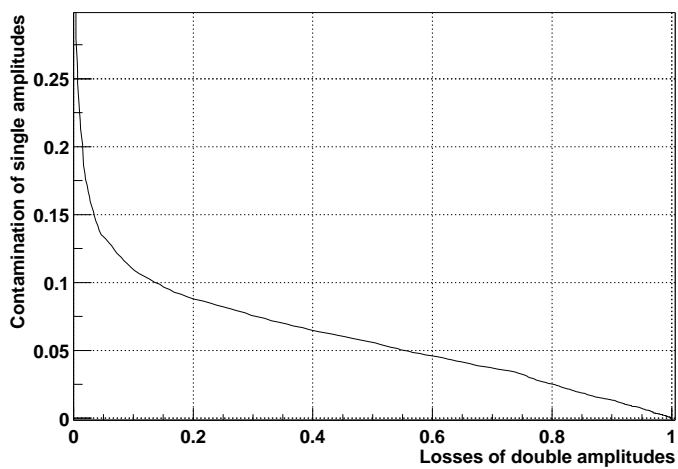}
\end{center}
\caption{Contamination of single ionisation amplitudes as a function
  of losses of double ionisation as obtained from the analysis of the
  spectra of Fig.~\ref{adcspectra}.}
\label{sdseparation}
\end{figure}

Scintillation light is detected by FEU-85 photomultipliers with 25~mm
diameter photocathodes.  Photomultipliers are assembled by 16 units
into a compact set, allowing independent replacement of each PM.
Photocathodes are in optical contact with the wide side of a light
guide instead of the traditional butt-end readout.  This improves the
light collection efficiency by about 50\%.  As the detector is highly
loaded by intense particle flux, the last 4 PM-dynodes are fed by an
additional power supply to ensure a constant PM amplification
throughout the spill.
 
Signal amplitude and time are digitised by LeCroy ADC\,4300B and
TDC\,3377 modules, respectively.  The time resolution of the IH
detector is better than 1~ns.  The typical response of one IH channel
to close particle pairs incident on one scintillating slab and to
single particle is shown in Fig.~\ref{adcspectra}.  If a threshold is
set to retain 90\% of the double ionisation signal from pairs, the
contamination from single particle amplitudes is less than 15\%
(Fig.~\ref{sdseparation}).

\section{Drift Chambers}

\subsection{General layout and characteristics}

The drift chamber system is used to perform particle tracking
downstream the dipole magnet.  The system is designed to sustain a
high particle fluency in the secondary channel, reaching
10~kHz/cm$^{2}$ at the innermost region.

A two-arm solution has been chosen, except for the first chamber which
is a single large module (DC-1) designed with two separated sensitive
areas $0.8\times 0.4$~m$^2$ each. This chamber provides 6 successive
measurements of the particle trajectory along the coordinates
X,Y,W,X,Y,W, where W is a stereo angle with inclination ${11.3}^{o}$
with respect to the X-coordinate.  DC-1 is instrumented with 800
electronic channels.

Each of the two arms consists of 3 chamber modules, of identical
design, measuring coordinates X,Y (DC-2), X,Y (DC-3) and X,Y,X,Y
(DC-4) following the direction of the outgoing particle.  Their
dimensions are $0.8 \times 0.4$~m$^2$ (DC-2), $1.12\times 0.4$~m$^2$
(DC-3), and $1.28 \times 0.4$~m$^2$ (DC-4). Both arms together contain
1216 electronic channels.

The distance between the center of the first half of DC1 and the
center of DC4 provides a lever-arm of 1.6~m along the average particle
path, having uniform spacing of chambers DC-2 and DC-3 along this
path. Characteristics of the drift chamber system are summarised in
Table~\ref{DC general}.

\begin{table}[htb]
\begin{center}
\caption{ General properties of the DC modules.}
\label{DC general}
\begin{tabular}{|c|c|c|c|}
\hline
Module   &  Sensitive   &    Measured      & Number of \\
type     & area, $cm^2$ & coordinate       & planes \\
\hline
DC-1     & $40\times80$ & $X$              & 2 \\
\cline{3-4}
         & left arm     & $Y$              & 2 \\
\cline{3-4}
         &              & $W$              & 2 \\
\cline{3-4}
         & $40\times80$ & $X$              & 2 \\
\cline{3-4}
         & right arm    & $Y$              & 2 \\
\cline{3-4}
         &              & $W$              & 2 \\
\hline
DC-2     & $40\times80$ & $X$              & 1 \\
\cline{3-4} 
         &              & $Y$              & 1 \\
\hline
DC-3     & $40\times112$ & $X$             & 1 \\
\cline{3-4} 
         &              & $Y$              & 1 \\
\hline
DC-4     & $40\times128$ & $X$             & 2 \\
\cline{3-4} 
         &              & $Y$              & 2 \\
\hline
\end{tabular}
\end{center}

\end{table}

\subsection{Drift chamber electrodes}
\label{Electrodes}

\begin{figure}[htb]
\begin{center}
  \includegraphics*[width=0.75\textwidth]{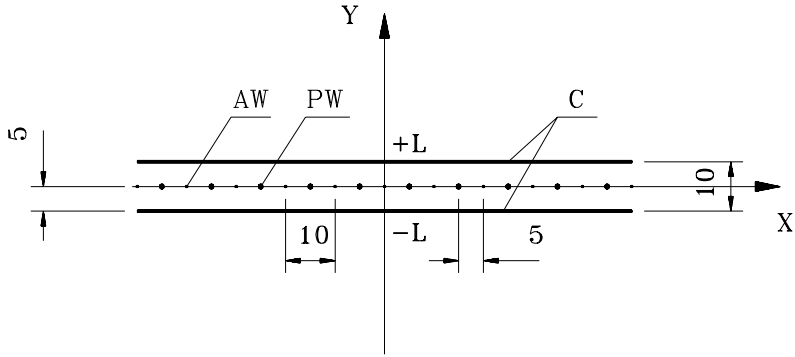}
\end{center}
\caption{ Schematic view of the wire chamber electrodes: $AW$ -- anode
  wires, $PW$ -- potential wires, $C$ -- cathode foils.  Dimensions
  are in mm.}
\label{dccell}
\end{figure}

A schematic drawing of the sensitive element is shown in
Fig.~\ref{dccell}. The anode wires pitch is 10~mm, the distance L
between the anode and cathode planes is 5~mm.  The cathode planes and
potential wires are at equal voltages. As seen in the figure, a
sensitive area, corresponding to each anode wire and limited by the
cathode planes and potential wires, has a square $(10\times10$ mm$^2)$
shape. In this case, with a suitable gas mixture, it is possible to
achieve a linear behaviour of the drift function, except in a small
region near the potential wire.

Cathode planes are made of $20~{\mu}$m thick carbon-coated mylar foils
with a surface resistivity of about $400~\Omega$ per square.  Such
cathode foils provide stable chamber operation due to a high work
function of the carbon coating and, being thin, add only small amount
of material along the particle path.

Anode and potential wires of $50~{\mu}$m and $100~{\mu}$m diameter,
respectively, are made of a copper-beryllium alloy. The rather large
diameter of the anode wires has been chosen in order to operate the
chambers at high current avalanche amplification mode.

\subsection{Chamber design}
\label{Design}

\begin{figure}[htb]
\begin{center}
  \includegraphics*[width=0.7\textwidth]{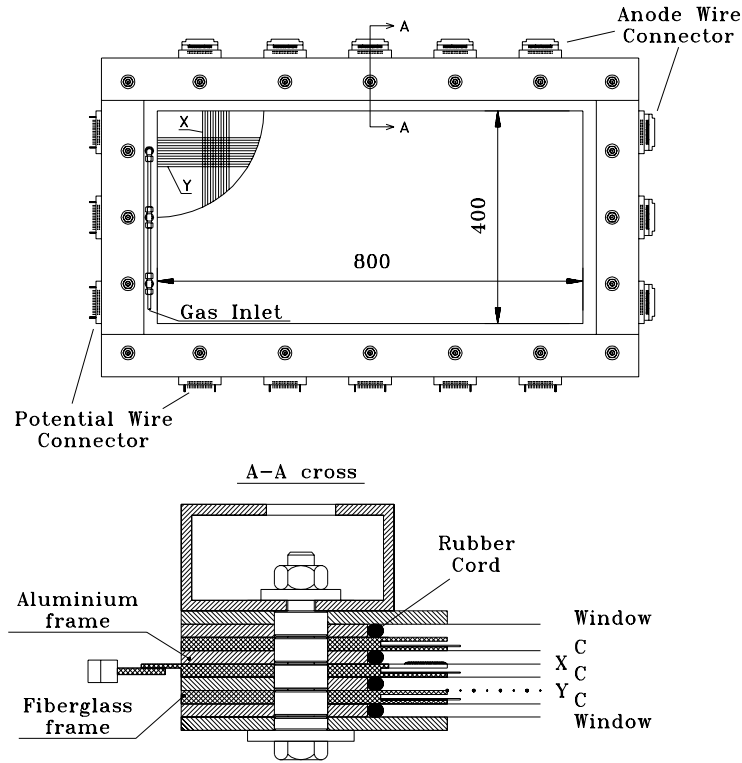}
\end{center}
\caption{ Design of the DC-2 module. Upper figure: general view.
  Lower figure: structure of the frame stack; X -- $X$-plane, Y --
  $Y$-plane, C -- cathode foils.}
\label{dc2}
\end{figure}

The chamber design is shown in Fig.~\ref{dc2} for the case of the DC-2
module.  The module is a stack of aluminium and fibreglass frames,
each of 5~mm thickness, fixed by screws.  The fibreglass frames are
the supports for the chamber electrodes (anode and potential wires and
cathode foils). The two outer aluminium frames in the stack are used
to fix the mylar window, and the inner ones are the spacers between
the fibreglass frames.  Rigidity of the module is enforced by
aluminium rectangular tubes screwed to the surface of the frame
package.

Gas tightness of the chamber module is provided by rubber o-rings
glued along the inner edges of the aluminium frames.  Within a module,
gas flows sequentially in the sub-volumes defined between cathode
foils, by means of holes drilled on opposite sides of the fibreglass
frames.

\begin{figure}[htbp]
\begin{center}
\includegraphics*[width=0.5\textwidth]{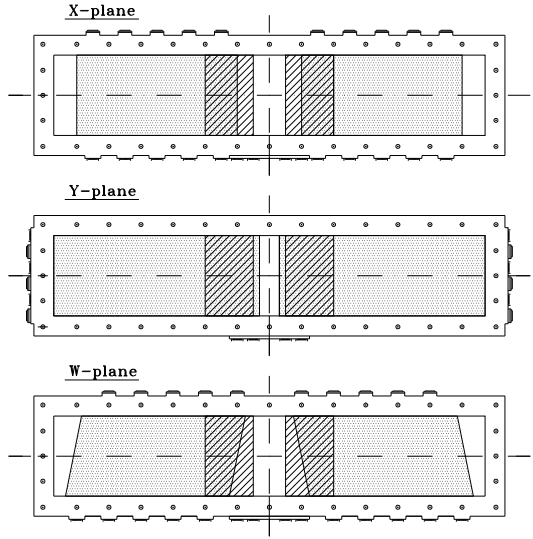}
\end{center}
\caption{ Schematic view of the $DC$-1 module. 
  Dotted areas show the sensitive regions of the $X$-, $Y$- and
  $W$-planes.  Hatched areas mark the zones of the cathode strips
  which allow to change the width of the insensitive area in the
  central region.}
\label{xyw}
\end{figure}

The design of module DC-1 differs from the one shown in
Fig.~\ref{dc2}.  The main difference, illustrated in Fig.~\ref{xyw},
consists in the fact that DC-1 comprises, in a single gas volume, two
sets of sensitive planes, placed symmetrically to the left and right
hand side of the spectrometer axis. The middle zone, which is strongly
irradiated by particles (mostly fast protons from target
fragmentation), is made insensitive to the particle flux. The limiting
edge of the sensitive zones, close to the axis, can be varied.  This
is possible by means of a stripped structure of the neighbouring
cathodes, which allows stepwise application of voltage.  This design
of the DC-1 module ensures little amount of material, by avoiding
frames in the small angle region.

\subsection{Chamber operation and performance}
\label{Operation}

The drift chambers operate in a high current avalanche mode.  This
mode is characterised by high pulse amplitude (about 1~mA), small
pulse width (20~ns), and stable operation due to an efficiency plateau
larger than 1~kV.  The single hit efficiency is above 96\% when the
particle flux is about 10~kHz/cm$^2$.  The employed gas mixture is
$Ar(\sim50\%)+iC_4H_{10}(\sim50\%)+H_2O(0.5\%)$, and the chamber
operation voltage is 3.85~kV.

\begin{figure}[htbp]
\begin{center}
  \includegraphics*[width=0.6\textwidth]{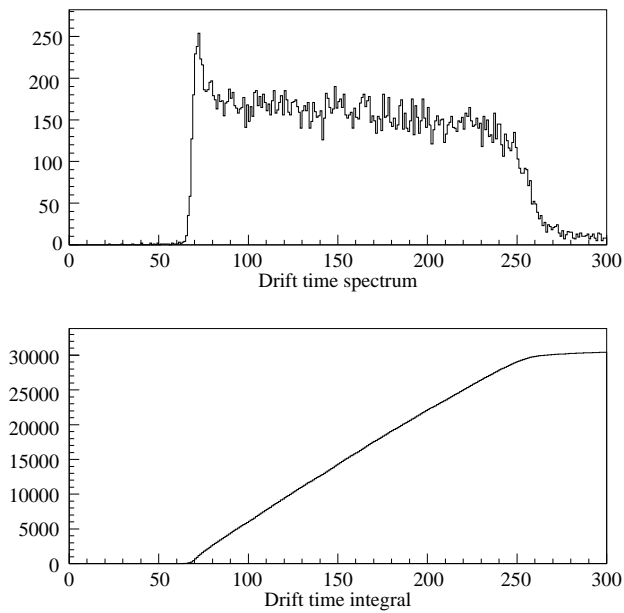}
\end{center}
\caption{ Distribution of the drift time (upper) and its integral 
spectrum (lower) for the $X4$-plane. Horizontal scale is in
TDC channels, bin width is $0.5~ns$.}
\label{drtime}
\end{figure}

A space-to-time relationship was extracted from the time spectrum and
its integral distribution shown in Fig.~\ref{drtime}, for a sample of
clean events with a small amount of background hits.  The integral
distribution has been parameterised by a second order polynomial of
the type:
$$
l=a_1\times{t^{\star}}+a_2\times{t^\star}^2.
$$
In this formula $t^\star=t_{TDC}-t_0-\delta{t}$, where $\delta{t}$
is the signal propagation time along the anode wire.

Study of the drift function parameters for different chamber planes at
different beam intensities shows good stability of the above relation.
For this reason the same drift function parameters ($a_1$ and $a_2$)
were used for all chamber planes during the off-line track
reconstruction procedure.

\begin{figure}[htbp]
\begin{center}
\includegraphics*[width=0.5\textwidth]{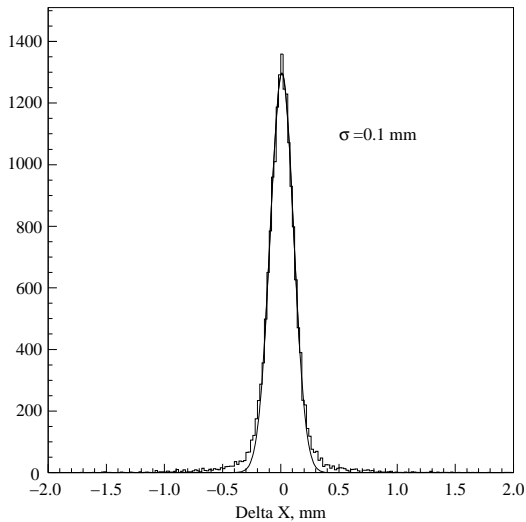}
\caption{ Distribution of differences between the measured and
predicted X-coordinate for one drift chamber plane ($X4$).}
\label{deltax}
\end{center}
\end{figure}

Coordinate resolution of the DC system is illustrated in
Fig.~\ref{deltax}, where the distribution of differences between the
predicted position and measured coordinates in one of the planes is
shown ($X4$-plane right arm).  The measured standard deviation,
$\sigma=100~{\mu}$m, is defined not only by the intrinsic chamber
plane resolution, but also by the accuracy of the predicted track
coordinates. Taking the latter into account the measured intrinsic
space resolution of one plane is better than $90~{\mu}$m.

Tracking efficiency of the drift chamber system as a whole is about
99\%, due to the fact that the requested number of hits per
reconstructed track is less than the total number of sensitive planes
crossed by a particle.

\subsection{Readout electronics}

The readout electronics of the drift chambers, which is a custom-made
system \cite{DCRS}, provides data readout into the data collection
memories and input to the trigger processor (see section ``Trigger
system'').

The sensitive wire signals are digitised in the 16-channel multi-hit
time-to-digital converter boards (TDC), which are plugged in the
connectors mounted onto the chamber frames. This solution results in
reduced number of electronic units, small number of cables and high
noise immunity.  The detection threshold in the TDC board can vary
from 0.05 to 2~mA, the maximum number of hits per channel is 16. Least
count of TDC is 0.5~ns, which corresponds to a drift distance of
25~$\mu$m at a constant drift velocity of $50~\mu$m/ns, well below the
chamber intrinsic space resolution.

A complete readout chain consists of TDC boards, bus drivers, readout
controller and VME memory. Up to 8 TDC boards can be connected to the
bus driver via common data and control buses, forming a segment of the
DC readout system. Similarly, up to 8 segments are connected to a
readout controller forming a readout branch.  The data of an event are
stored in local data buffers until the higher level trigger decision
is issued.  If the event is accepted, the data are serially
transferred via the readout controller to the VME memories.  To read
out all the DC data 3 readout branches are used.  The accepted data
are transferred to the VME buffers within 5~$\mu$s on average.  The DC
readout is fast enough compared to the global readout time of the
experiment.

For trigger purposes the DC readout system is equipped with fast data
ports which directly transmit the hit wire numbers to a trigger track
processor.  This provides a minimum access time to the data, thus
reducing the latency of the trigger system. Transmission to the track
processor is performed in parallel.
        
\section{The Vertical Hodoscopes (Time-of-Flight Detector)}

The vertical hodoscope (VH) consists of an array of vertical
scintillating slabs placed downstream the DC system. The VH system,
together with the horizontal hodoscope, provides fast coincidence
signals between the spectrometer arms necessary for the first level
trigger.  It is also used, in correlation with other detectors, in the
definition of dedicated triggers for calibration purposes and of a
higher level trigger for the selection of low $Q$ events (see trigger
section later).

\begin{figure}[htb]  
\begin{center}
  \includegraphics*[width=11cm,height=10cm]{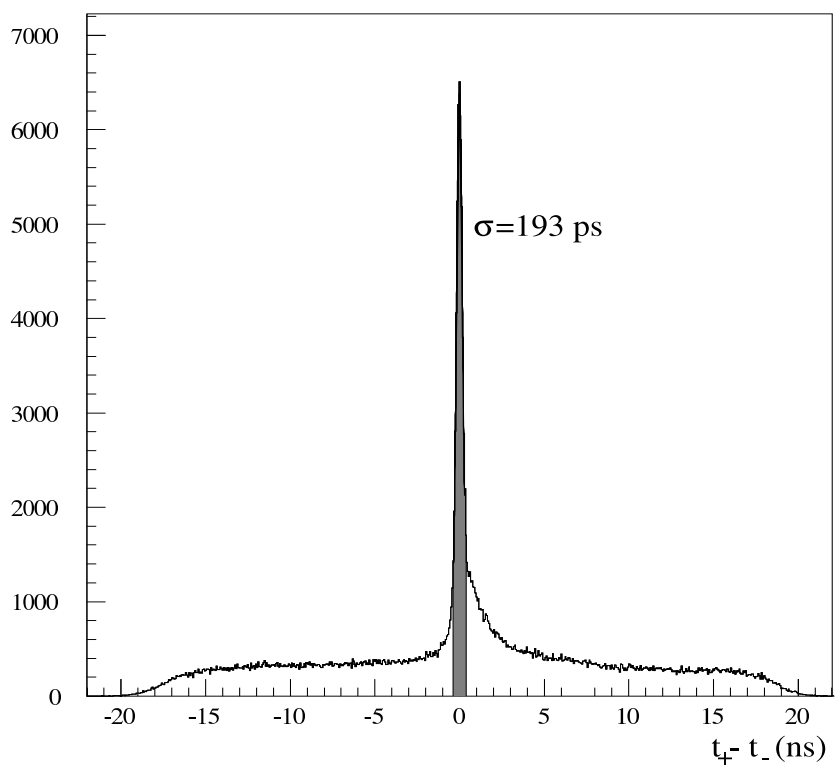}
  \caption{\label{tof1}  Time 
    difference between charged positive and negative particles
    obtained from standard hadron trigger data. The central peak has a
    gaussian width of 193~ps, and the shaded area represents a $ 2
    \sigma $ cut used to select prompt $\pi \pi$ events. The flat
    background is originated from accidental pairs, not belonging to
    the same beam interaction. Note the shoulder on the right-hand
    side of the peak, due to $\pi^- p$ prompt pairs.}
  \end{center}
\end{figure}

A key function of this detector, which motivated a special design, is
to provide a very accurate time definition of pion pairs originated
from the same proton interaction (prompt pairs), in order to perform a
clean separation (in off-line analysis) with respect to pairs in which
the pions are produced at different times (accidental pairs).
 
Used as a time-of-flight detector it allows to identify $p \pi^-$
pairs in prompt events, as they might constitute a significant source
of background to the $\pi^+ \pi^-$ signal.

The VH detector consists of two identical telescopes matching the
acceptance of the drift chamber system previously described.  Each
telescope contains an array of 18 vertical scintillation counters.
The scintillating material is BICRON BC420 and the slab dimensions are
40~cm length, 7~cm width and 2.2~cm thickness.  Scintillation light is
collected at both ends by two 12-dynode Hamamatsu R1828-01
photomultipliers (chosen because of their small transition time
spread) coupled to fish-tail light guides.  Voltage dividers were
designed to provide sufficient bleeder current at particle rates up to
2$\times {10}^6$ Hz, while preventing any degradation of the time
resolution.

The front-end electronics was designed to minimise the time jitter.
This is achieved by using LeCroy L3420 constant fraction
discriminators, followed by CAEN C561 meantimers to provide a position
independent time measurement.  Hodoscope signals are delayed and
transferred to the data acquisition system using shielded twisted-pair
cables.  Time digitisation is performed by LeCroy 4303 time-to-FERA
Converter followed by ADC 4300B. The least count of this ensemble is
62~ps.

The VH single-hit detection efficiency is 99.5\% for the positive, and
98.8\% for the negative hodoscope arms. In Fig.~\ref{tof1} the
distribution of the time difference between positive and negative
pions in the spectrometer is shown.  The observed ratio between prompt
and accidental pairs in the 2$\sigma$ cut region around the peak is
about 16.

\begin{figure}[htb]
\begin{center}
  \includegraphics*[width=12cm,height=8cm]{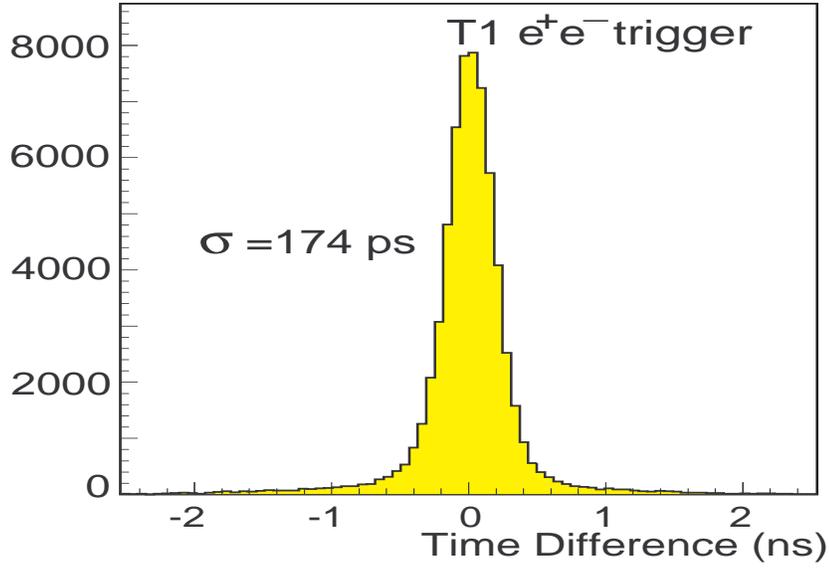}
\end{center}
\caption{Time difference spectrum for $e^+e^-$ pairs detected by the vertical 
  hodoscopes, after path length correction. Data come from a sample of
  $e^+e^-$ calibration triggers.}
\label{e+e-}
\end{figure}

The overall time resolution of the system has been measured with $e^+
e^-$ pairs to be 127~ps per counter \cite{tofnim}, which corresponds
to 174~ps accuracy for the time difference between positive and
negative arms (time-of-flight resolution).  The latter is shown in
Fig.~\ref{e+e-}.

\begin{figure}[htb]   
  \begin{center}
    \includegraphics*[width=10cm]{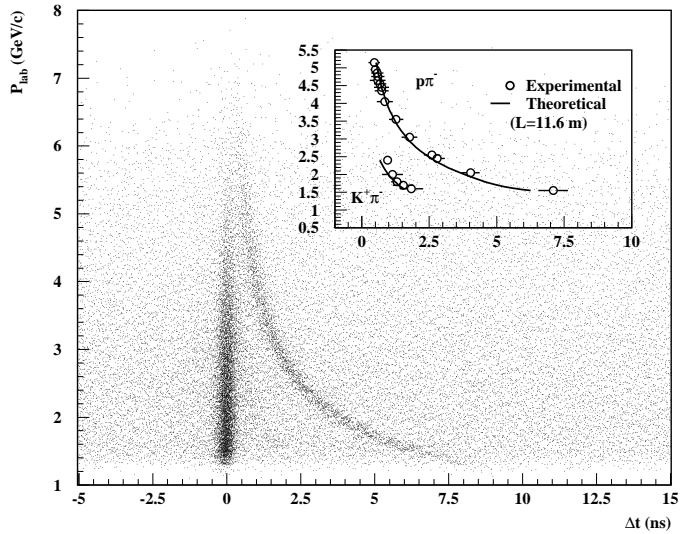}
  \end{center}
\caption{\label{tof3}  Correlation between the
  measured momentum of the positive particle and the VH time
  difference between the positive and negative spectrometer arm,
  taking into account the correction for the difference in path
  length. The accumulation bands correspond to $\pi^-\pi^+$ (vertical
  band) and $\pi^- p$ (curved band) pairs.  A small cluster of $\pi^-
  K^+$ pairs is also visible in the intermediate region.}
\end{figure}

The dedicated $e^+ e^-$ calibration trigger selects $e^+ e^-$ pairs
from $\gamma$ conversions and Dalitz decays of $\pi^0$ which are
almost synchronous in time, as the time of flight of $e^+ e^-$ pairs
is momentum independent in the available setup range of momenta.

This timing capability allows to separate $\pi^+ \pi^-$ from $\pi^- p$
pairs in the momentum range from 1 to 5~GeV/$c$, and from $\pi K$
pairs in the range from 1 to 2.5~GeV/$c$, as illustrated in
Fig.~\ref{tof3}.

\section{The Horizontal Hodoscopes}

The horizontal hodoscope system (HH) is also separated into two arms,
each covering an area of $40\times130$~cm$^2$. The HH participates,
together with the vertical hodoscope, to the definition of the first
level trigger. Its response is used to apply a coplanarity criterion
to track pairs hitting both detector arms. This trigger requirement
selects oppositely charged particles with relative vertical
displacement, $\Delta y$, less than 7.5~cm.

\begin{figure}
  \begin{center}
    \includegraphics*[height=7cm,bb=120 45 450 355]{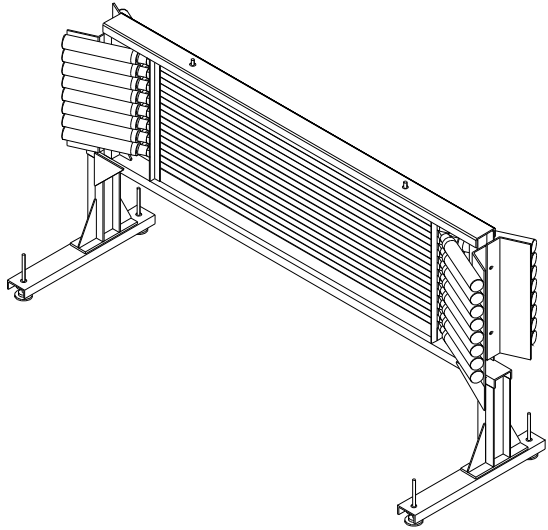}
    \caption{\label{hh}  General layout of the horizontal hodoscopes.} 
  \end{center} 
\end{figure}

Each hodoscope consists of 16 horizontal extruded scintillating slabs
of dimensions $130\times2.5$~cm$^2$, with a thickness of 2.5~cm.  Both
ends of each slab are coupled to specially shaped light-guides as
illustrated in Fig.~\ref{hh}.

Photoconversion is done by Philips XP2008 photomultipliers, equipped,
as for the VH, with a voltage divider allowing high counting rate
capability. The front-end electronics system contains the same
elements as those described for the vertical hodoscopes.  The single
hit detection efficiency of HH is greater than 96.6 \% on both arms,
and the time resolution is 320~ps.

\section{The Cherenkov Counters}

This detector is essential for rejection of the main background of
electron-positron pairs from photon conversion, Dalitz pairs, and to a
minor extent from resonance decays.  It is used in the first level
$\pi^+ \pi^-$ main trigger and in the calibration trigger to select
$e^+ e^-$ pairs.

\begin{figure}[htb]
\begin{center}
  \includegraphics*[width=12cm,height=8cm]{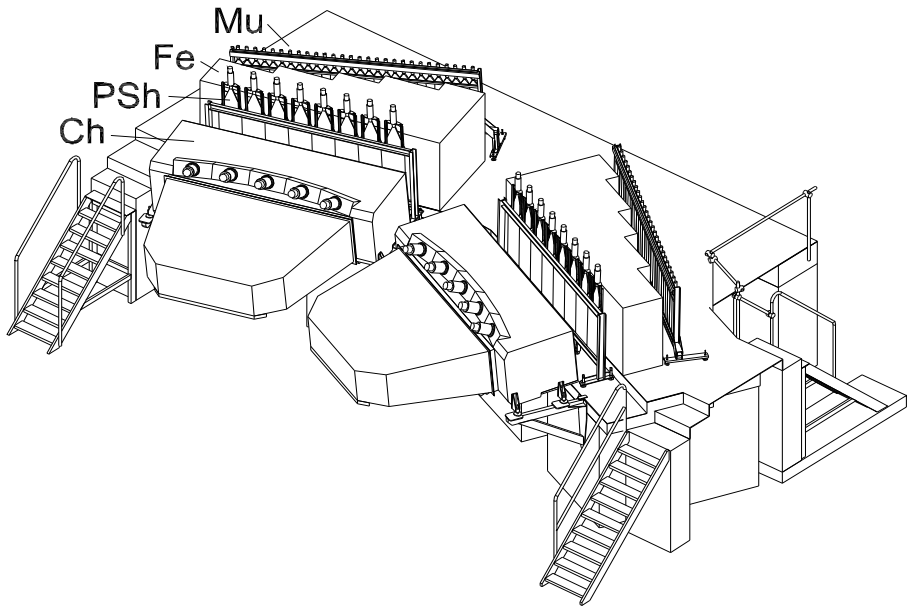}
\caption{\label{ch1}  
  The far end part of the DIRAC setup, comprising threshold Cherenkov
  counters (Ch), preshower detector (PSh), iron absorber (Fe) and muon
  counters (Mu).}
\end{center} 
\end{figure}

It is structured in two identical threshold Cherenkov counters
\cite{cheren}, each covering one spectrometer arm (see
Fig.~\ref{ch1}).

The gas radiator is enclosed in a volume defined by the entrance and
exit windows, with dimensions $143\times 56$~cm$^2$ and $336\times
96$~cm$^2$, respectively.  The chosen radiator is $N_2$ at normal
temperature and pressure ($\theta_{\check{C}}={1.4}^\circ$) and the
counter length is 285~cm.

\begin{figure}
\begin{center}
  \includegraphics*[height=6cm,width=\textwidth,bb=0 135 550
  405]{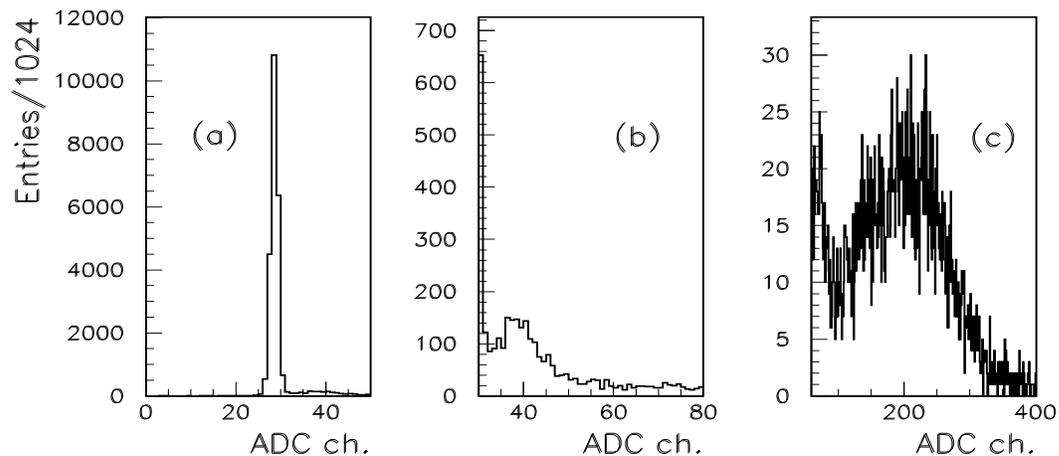}
\caption{\label{ch2}  ADC spectrum from one Cherenkov
  photomultiplier: (a) spectrum from pions (practically equal to the
  ADC pedestal distribution), (b) amplitude signal from single
  photoelectron, (c) spectrum from electrons.}
\end{center} 
\end{figure}

\begin{figure}
\begin{center}
  \includegraphics*[height=0.4\textwidth,width=0.8\textwidth,bb=15 90
  525 441]{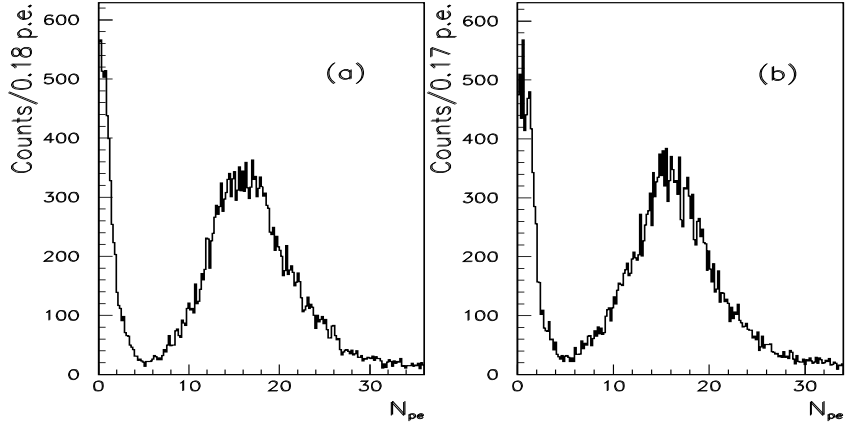}
\caption{\label{ch3}  Distribution of the number of photoelectrons
  detected from the (a) positive and (b) negative Cherenkov detector
  arms.}
\end{center} 
\end{figure}

Each counter is equipped with 20 mirrors and 10 photomultipliers on
two rows. Cherenkov light reflected by pairs of adjacent mirrors is
focused onto the same photomultiplier (Hamamatsu R1587, with 130~mm
UV-glass window).  Mirrors and PM arrays are placed along
circumferences with radii 630~cm and 570~cm, respectively, centred at
the magnet middle point. Mirrors are spherically deformed rectangles,
with average dimensions $30\times35$~cm$^2$ and 6~mm thickness.

The analog signals from individual PM are fed into two custom-made
summing modules, one per counter (10 channels input). The output of
the summing module is a linear sum with fan-out of 2, plus the 10
input signals.  At the module output the linear analog sum is
attenuated by a factor of 3 with respect to individual channels.  One
analog sum is discriminated (LeCroy 4413 leading edge discriminators)
and used for trigger purposes, whereas the individual PM signals and
the other sum are fed, after a delay line, into LeCroy 4300B ADC
units.  Individual PM ADC spectra are used to perform overall
amplitude alignment on the basis of maximum ratio between mean and
r.m.s., in the PM voltage range $2400 \pm 200$~V.

A single photoelectron peak is clearly observed in all channels and it
is shown in Fig.~\ref{ch2}(b). The single photoelectron spectrum was
used to cross-check the conversion factor from ADC counts to number of
photoelectrons ($N_{pe}$), which is obtained from the analysis of the
summed amplitude spectra. Fig.~\ref{ch3} shows the latter as a
function of $N_{pe}$ for the positive (a) and the negative (b)
spectrometer arms.  The mean values are $N_{pe}$=16.2 and
$N_{pe}$=16.4, respectively. From these values we infer that both
counters have an efficiency greater than 99.8 \% when operated at a
threshold slightly less than 2 photoelectrons. The pion contamination
above the detection threshold is estimated to be less than 1.5 \%.
Such contamination arises from pions with momenta above the Cherenkov
threshold and from accidental coincidences occurring within the
trigger time-window.

A general picture of the installed detectors described so far, looking
into the magnet from a distance close to that of the Cherenkov
counter, can be appreciated in Figure \ref{down_photo}.

\begin{figure}
\begin{center}
\caption{General picture of the spectrometer installed
  downstream the magnet, taken from a reference plane close to the
  Cherenkov counters, which can be appreciated in the bottom part of
  the picture.  The magnet can be seen at the far end, and in between
  we see the vertical and horizontal hodoscope photomultiplier layout.
  The drift chamber system is installed behind it. }
\label{down_photo}
\end{center} 
\end{figure}

\section{The Preshower Detector}

The purpose of the Preshower detector (PSH) is two-fold: it provides
additional electron/pion separation power in the off-line analysis and
is used in the trigger generating logic as well.

\begin{figure}[htb]
\begin{center}
  \includegraphics*[width=12cm]{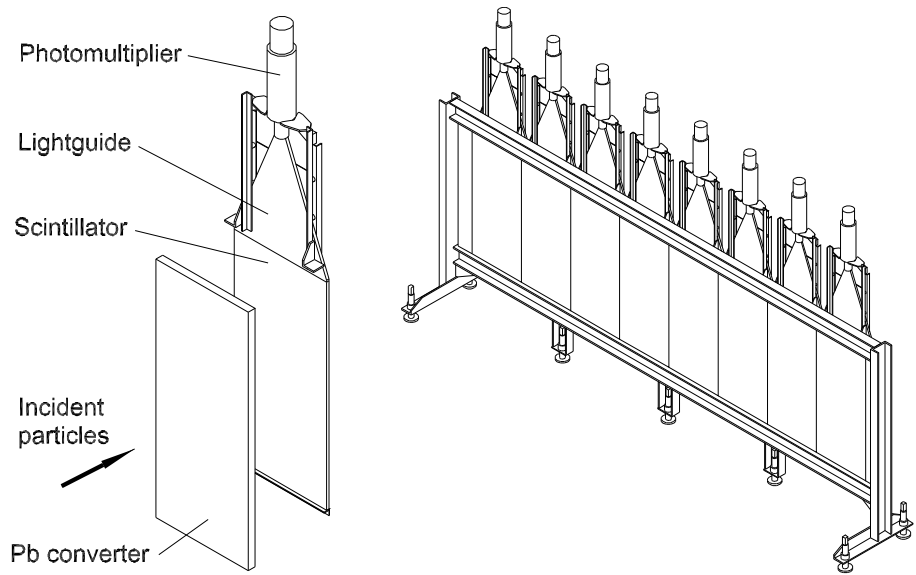}
\end{center}
\caption{PSH element and PSH array on one spectrometer arm.}
\label{prs}
\end{figure}

The PSH is based on an array of lead converters followed by
scintillation detectors \cite{pentia}. Electrons (positrons) initiate
in the converters electromagnetic showers which are sampled in the
scintillation counters while pions behave mainly as minimum ionising
particles.  For trigger purposes, the signal has to be produced
whether a pion or an electron cross a PSH counter.

The PSH consists of 16 detector elements placed symmetrically in two
arms, as seen in Fig.~\ref{ch1}. Each element has a Pb converter and a
scintillation counter, Fig.~\ref{prs}.  The converters of the two
outermost elements of each arm (low momentum region) are 10~mm thick,
whereas the rest are 25~mm thick (around 2 and 5 units of radiation
length, respectively).  The scintillator used is BICRON type BC-408,
with slab dimensions $35\times75$~cm$^2$ and 1~cm thickness.  The
scintillation light is transmitted to photomultipliers EMI 9954-B,
placed at one end only, by 10~mm thick Plexiglas light-guides ending
with Plexiglas cylinders to match with the PM photocathodes. Since the
maximum particle flux on each PSH element is as high as 2~MHz, an
additional booster power supply is used to feed the last PM dynodes.

The detector signals are linearly split into two branches, one used
for trigger purposes and another for ADC analysis.  In the former, a
leading edge LeCroy 4416 discriminator is used with a threshold
corresponding to efficient detection of minimum ionising particles. In
the latter the signals are fed into 20~dB attenuators in order to fit
the electron signals into the dynamical range of the LeCroy 4300B ADC.

\begin{figure}[htb]
\begin{center}
  \includegraphics*[height=8cm,bb=0 18 522 512]{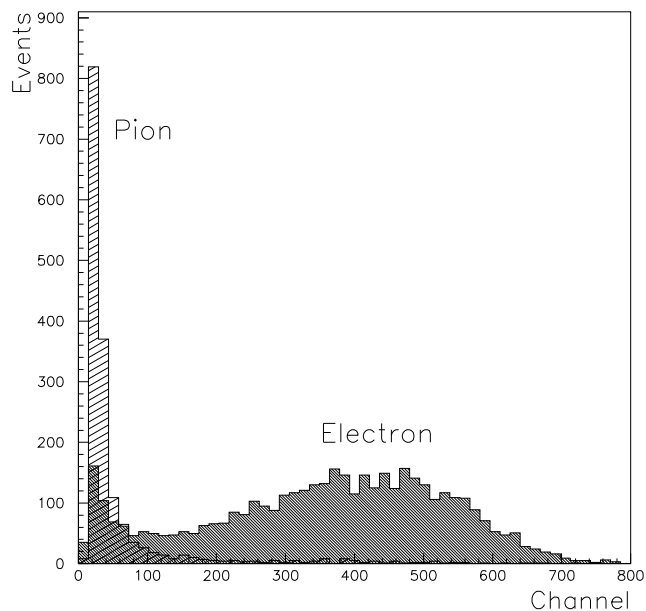}
\end{center}
\caption{Pulse-height spectra for pions and electrons
  in one element of PSH.}
\label{pshspectrum}
\end{figure}

The single arm detector efficiency is 99.5\% for pions.
Fig.~\ref{pshspectrum} shows the pulse-height spectra from one element
of the PSH for pions and electrons selected by the trigger system
(this selection is based on the Cherenkov detector response).

As seen from Fig.~\ref{pshspectrum}, the pion spectrum has a tail
(originating from nuclear interaction of pions in the converter)
extending to the electron amplitude region.  The off-line study of the
$e/ \pi$ separation efficiency of the PSH showed that rejection of
electrons reaches 85\% with less than 5\% losses of pions. As only a
very small fraction of electrons escapes the on-line identification in
the Cherenkov counters, the combined use of the Cherenkov and PSH
detectors provides almost 100\% electron rejection power at the
off-line analysis stage.

\section{Muon detector}

Admixture of muons in the $\pi\pi$ events can be a serious source of
background. For this reason a muon detection system is implemented to
provide efficient muon tagging. Muons come almost entirely from pion
decays with a small admixture from other decays and direct $ \mu^+
\mu^-$ pair production.

The muon detector consists of scintillation counters placed behind a
thick iron absorber which almost entirely absorbs hadrons and related
hadronic showers. This detector is placed at the downstream end of the
DIRAC apparatus, few meters from the intense primary proton beam dump.
As a result, the muon scintillation counters may undergo a high flux
of background radiation from the beam dump area.  This has required a
special design of the counter arrays and electronics and has prevented
from using muon information during on-line data reduction.
 
The counters are located behind iron absorber blocks with thickness
ranging from 60 to 140~cm (see Fig.~\ref{ch1}).  The thickness is
larger in the region close to the spectrometer symmetry axis, in order
to compensate for the harder pion momentum spectrum.  A double layer
structure has been envisaged for the counters, each layer on each arm
consisting of 28 counters with equal scintillating slabs of
$75\times12$~cm$^2$ front area and 0.5~cm thickness.  The muon
detector data are read out only if simultaneous signals from a pair of
corresponding counters in the two layers are detected. This
essentially reduces the background counting rate induced by the
neutron flux from the beam dump.

\begin{figure}[htb]
\begin{center}
  \includegraphics*[width=7cm]{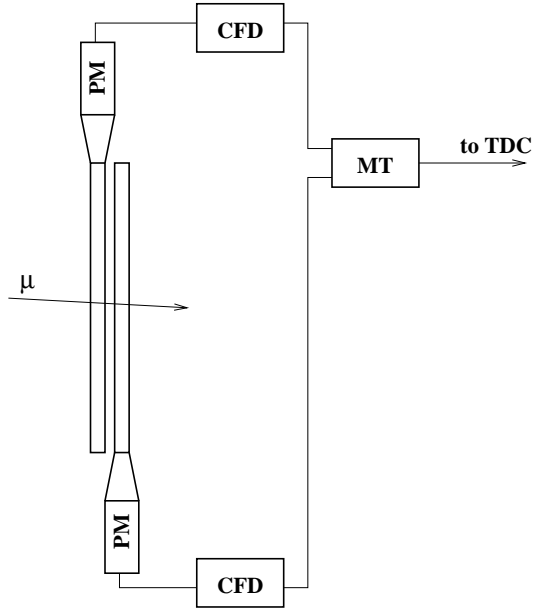}
\caption{\label{muonreadout} Readout scheme of the muon detector element.} 
\end{center} 
\end{figure}
A special readout architecture is realised to compromise between a
cost-saving solution and the need to achieve a reasonably high time
resolution. Scintillation light is detected by 25~mm diameter
bialkaline FEU-85 photomultipliers placed at one scintillator end, in
the two layers at opposite ends, as shown in Fig.~\ref{muonreadout}.
Signals from a pair of counters are fed into constant fraction
discriminators (CFD) followed by meantimer (MT). CAEN modules C808 and
C561 are used, respectively, for this purpose. In such a scheme the
output signal is generated only if both counters are hit, and correct
timing occurs only if the same particle crosses both counters.

\begin{figure}
\begin{center}
  \includegraphics*[height=6.5cm,bb=70 49 510 390]{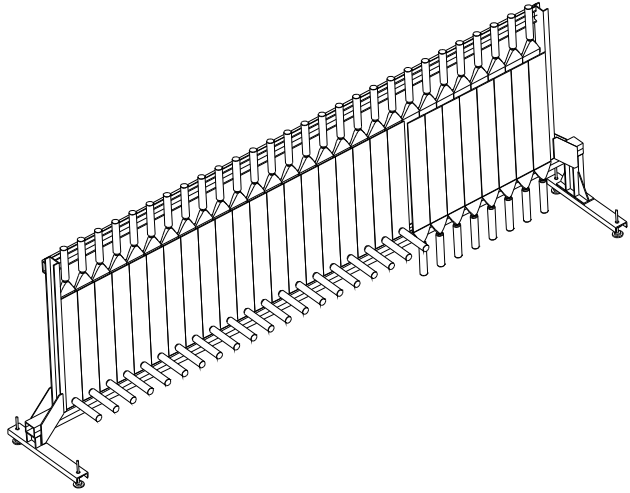}
\caption{\label{muon1}  Schematic layout of muon counters on
  their support structure, indicating light guides and
  photomultipliers.}
\end{center} 
\end{figure}

Fish-tail light guides are used to couple the PM photocathodes to the
scintillators, except when this is impeded by the presence of the
concrete floor (see Fig.~\ref{muon1}).

\begin{figure}[htb]
\begin{center}
  \includegraphics*[width=8cm]{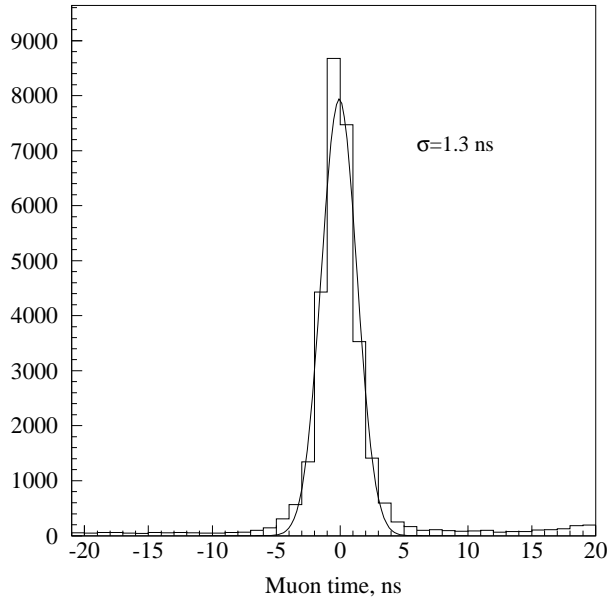}
\caption{\label{muontime} Time difference  
  between the signals of the muon detector and the vertical
  hodoscope.}
\end{center}
\end{figure}
In this case (for 20 counters of the second layer in each arm),
photomultipliers are directly coupled to the scintillating slabs,
which are then made twice thicker to compensate for the loss of light
yield.

In Fig.~\ref{muontime} the measured time difference between the
signals of the muon detector and the vertical hodoscope is shown for
an individual muon element of the positive arm. A global time
resolution of 1.3~ns is observed, with a very small background level.

The way to suppress muons at the trigger level would be to include the
muon counter signals into the anti-coincidence trigger logic.  Because
of the high background load to the muon detector, one might
dangerously suppress useful pion events if they happen to occur
on-time with background signals in the muon detector.  That is why the
option to use off-line the muon detector information has been chosen.
In the off-line analysis only the events with muon counter hits
correlated in time with those of other detectors are tagged as
``muon'' events and thus rejected \cite{brekhov}. From the analysis of
experimental data we have inferred that the fraction of such events,
containing at least one muon, is about 10\% \cite{armando_mu}. Such
muon-events originate to a large extent ($\sim$ 80\%) from
$\pi^{\pm}$-decays in the path between the DC and the muon counter
systems (decays upstream the DC are mostly suppressed by the trigger
system and thus contribute less to the collected event sample).

\section{Trigger system}

The trigger system was designed to provide a reduction of the event
rate to a level acceptable to the data acquisition system which is
around 2000 events/spill. Pion pairs are produced in the target mainly
in a free state with a wide distribution over their relative momentum
$Q$, whereas atomic pairs from $A_{2 \pi}$ disintegration have very
low $Q$, typically below 3~MeV/$c$. The on-line data selection rejects
events with pion pairs having $Q_L>30$~MeV/$c$ or $Q_x>3$~MeV/$c$ or
$Q_y>10$~MeV/$c$, keeping at the same time high efficiency for
detection of pairs with $Q$ components below these values, ($Q_L, Q_x$
and $Q_y$ are longitudinal and transversal components of the relative
momentum, respectively.)
 
A multilevel trigger is used in DIRAC \cite{mltrigger}. It comprises a
simple and fast first level trigger and higher level trigger
processors which apply selection criteria to different components of
the relative momentum of pion pairs.

Due to the requirements of the data analysis procedure, the on-line
selection of only time correlated (prompt) pion pairs, originating
from a single proton-target interaction and detected simultaneously by
both spectrometer arms, is not enough. In addition, a large number of
uncorrelated, accidental, pion pairs is also necessary. These
accidental pairs are used in the off-line analysis to describe the
relative momentum distribution of free (non-atomic) pion pairs without
Coulomb interaction in the final state.  Therefore, the trigger system
should apply very similar selection criteria to prompt and accidental
events, within a preselected coincidence time window centred around
the peak of prompt events.  The statistical error of the $A_{2 \pi}$
lifetime measurement depends on the number of both prompt and
accidental detected pairs. In standard experimental conditions, the
optimal ratio of prompt to accidental events is obtained using a 40~ns
wide coincidence time window between the times measured in the left
(VH1) and right (VH2) vertical hodoscopes.

Since 1999, when the experiment has started, the trigger architecture
was upgraded several times to achieve a larger reduction of the
background event rate (prompt and accidental pairs with large values
of $Q$). In the present article we briefly describe the most recent
version \footnote{ Before 2001 the trigger system included T2 and T3
  stages following the first level trigger (T1). T2 selected particle
  pairs with a small $\Delta x$ distance in the upstream spectrometer
  region (for rejection of high $Q_x$) using the data from SFD and IH.
  T3 analysed hit patterns of the upstream IH and downstream VH
  detectors imposing selection criteria to $Q_L$ values. A detailed
  description of T2 and T3 is given in \cite{mltrigger} and \cite{T3}.
  With the implementation of a new architecture of the trigger system
  these trigger stages were removed.}.

\begin{figure}[htb]
\begin{center}
  \includegraphics*[width=10cm,height=5cm]{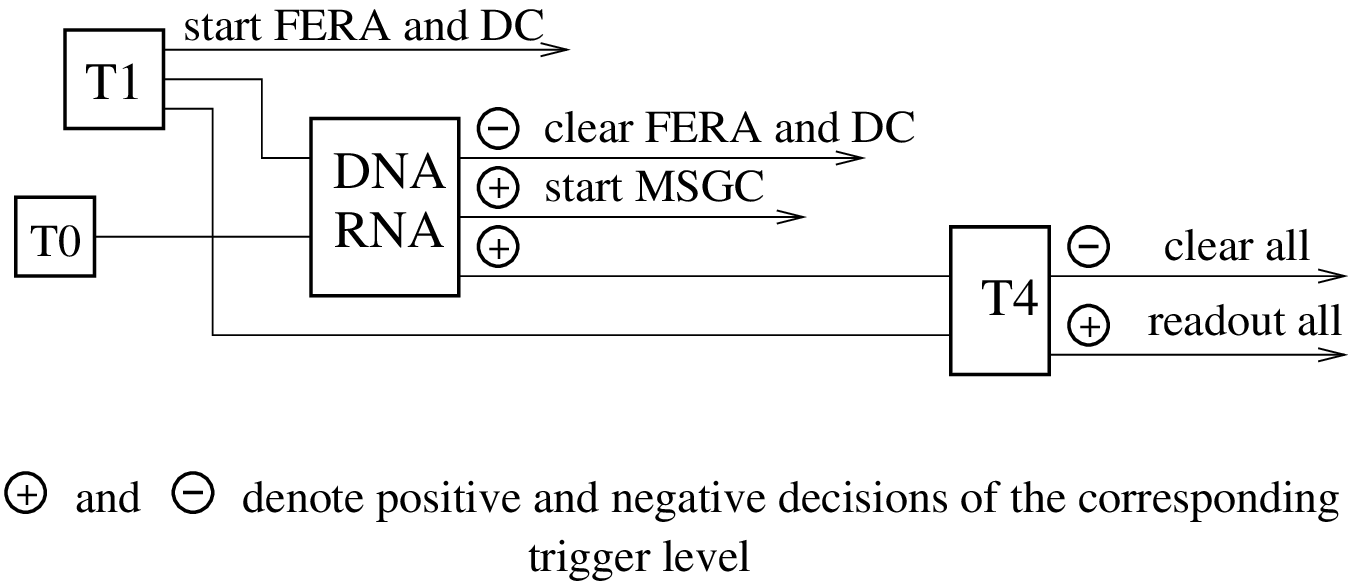}
\end{center}
\caption{General block diagram of the DIRAC multilevel trigger.}
\label{structure}
\end{figure}

A block diagram of the trigger architecture is presented in
Fig.~\ref{structure}.  The first level trigger T1 starts digitisation
of the detector signals in the data acquisition (DAQ) modules (ADC,
TDC, etc.). At the next level the neural network trigger DNA/RNA
(DIRAC Neural Atomic and Revised Neural Atomic trigger) rejects the
events with high $Q$ values.  At the last stage, a powerful drift
chamber trigger processor T4 imposes additional constraints to the
relative momentum and takes the final decision to accept or to reject
the event.

In addition to the main trigger designed to detect pionic atoms,
several calibration triggers are run in parallel.  They are applied to
DAQ directly, without suppression by higher level trigger stages but
with appropriate prescaling factors.

\subsection{First level trigger (T1)}

The first level trigger (a detailed description is given in \cite{T1})
fulfils the following tasks:
\begin{itemize}
\item[---] Selects events with signals in both detector arms
  downstream the magnet.
\item[---] Classifies the particle in each arm as $\pi^{\pm}$ or
  $e^{\pm}$ depending on the presence of the Cherenkov counter signal.
  Protons, kaons and muons are equally included in the ``pion'' class,
  their identification is performed in the off-line analysis.
\item[---] Arranges the coincidences between the signals detected in
  the two arms.  The width of the coincidence time window defines the
  ratio between yields of prompt and accidental events
  in the collected data.
\item[---] Applies a coplanarity criterion to particle pairs: the
  difference between the hit slab numbers in the horizontal hodoscopes
  in the two arms (HH1 and HH2) should be $\leq 2$. This criterion
  forces a selection on the $Q_y$ component of the relative momentum
  and provides a rate reduction by a factor of 2.
\item[---] Selects in parallel events from several physics processes
  needed for the setup calibration: $e^+e^-$ pairs, $\Lambda
  \rightarrow p + \pi^-$ decays, $K^{\pm}$ decays to three charged
  pions.
\end{itemize}
The physics and calibration trigger signals pass through the mask
register and, after proper prescaling, are combined with an OR
function. Any trigger type can be enabled or disabled by proper
programming of the mask register.  Independent prescaling of each
sub-trigger channel allows to adjust their relative rate with respect
to the rate of the main trigger. A specific trigger mark is recorded
for every event to allow sorting the data by trigger type during
off-line analysis and on-line monitoring.

All T1 modules are ECL line programmable multichannel CAMAC units.
Most of them are commercial modules, except for the dedicated
coplanarity processor which has been custom-developed at JINR.
Meantimer units are used in all VH and HH channels to remove the
dependence of the time measurement on the hit location, thus reducing
the total trigger time jitter.

\subsection{Neural network trigger (DNA/RNA)}

The DNA/RNA trigger \cite{DNA} is a processing system using a neural
network algorithm. Its hardware is based on the custom-built version
of the neural trigger used in the CPLEAR experiment
\cite{DNAhardware}.

\begin{figure}[htb]
\begin{center}
  \includegraphics*[width=11cm]{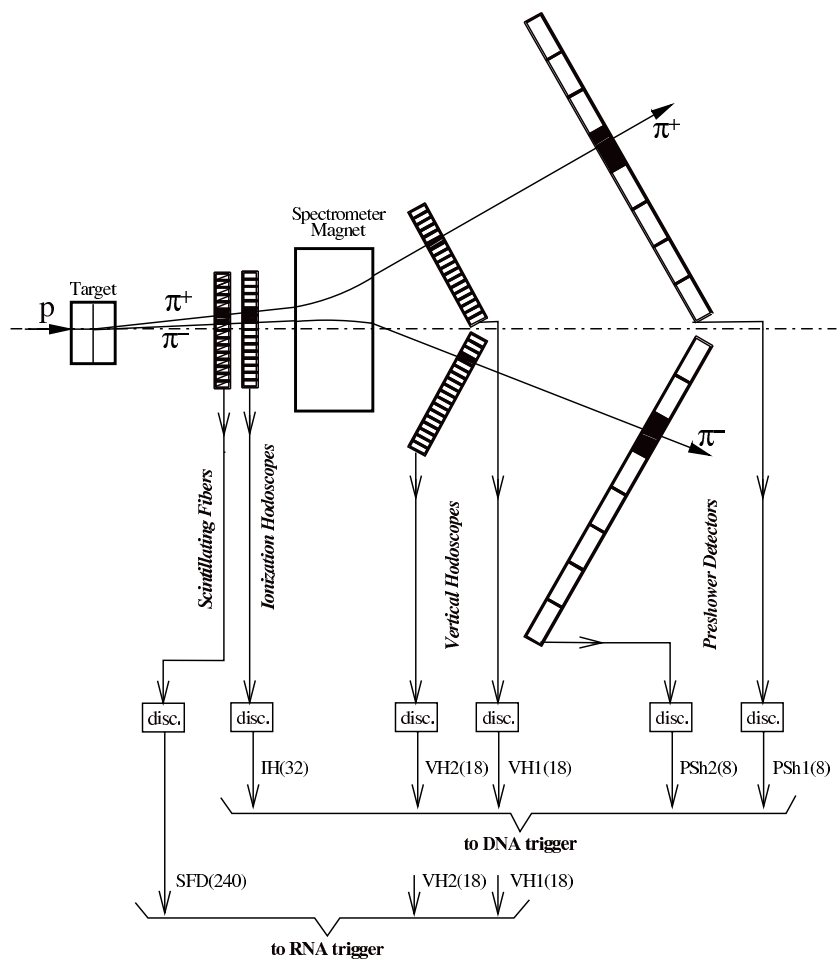}
\end{center}
\caption{DIRAC detectors used for the neural network triggers DNA and
  RNA. Numbers of signal channels from each detector are given in
  parentheses.}
\label{rna}
\end{figure}

DNA/RNA receives (see Fig.~\ref{rna}) the hit patterns from the
vertical hodoscopes VH1, VH2 and the X-planes of the upstream
detectors: the ionisation hodoscope (IH) and the scintillating fibre
detector (SFD). For low $Q$ events the hits in these detectors are
correlated.

The neural network was trained to select particle pairs with low
relative momenta: $Q_x<3$~MeV/$c$, $Q_y<10$~MeV/$c$ and
$Q_L<30$~MeV/$c$.  The events which do not satisfy any of those
conditions are considered ``bad'' and rejected.

The DNA/RNA logic is started by a fast pretrigger, T0, and in 250~ns
evaluates an event. Rate reduction by a factor of 2 with respect to T1
is achieved with DNA/RNA.

\subsection{Drift chamber processor (T4)}

T4 is the final trigger stage. T4 processor reconstructs straight
tracks in the X-projection of the drift chambers and analyses them to
determine the value of the relative momentum (the algorithm is
described in \cite{mltrigger}).

\vspace*{1cm}
\begin{figure}[htb]
\begin{center}
  \includegraphics*[width=11cm]{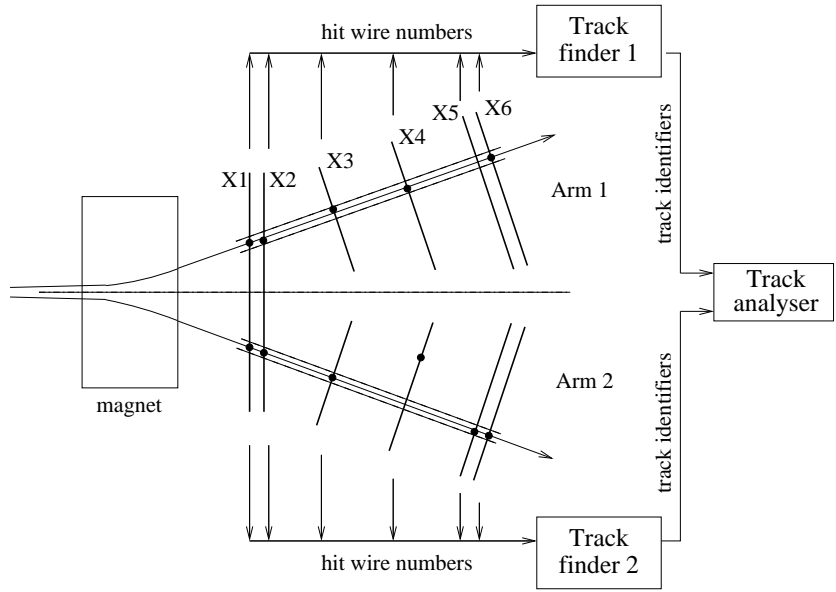}
\end{center}
\caption{T4 operation block diagram. Only the drift chamber X-planes
  involved in T4 are shown. }
\label{T4}
\end{figure}

The block diagram of the T4 operation is shown in Fig.~\ref{T4}.  The
drift chamber processor includes two stages: the track finder and the
track analyser. The track finder (an identical processor is used for
each arm) receives the numbers of the hit wires from all drift chamber
X-planes. Drift time values are not used in the T4 logic. A unique
number, ``track identifier'', which contains the encoded numbers of
the hit wires, is associated to the found track.

If tracks are found in both arms, the track analyser continues the
event evaluation. The track analyser receives the track identifiers
from both arms and compares them with the content of a look-up memory
table which contains all possible combinations of track identifiers
for pion pairs with $Q_L<30$~MeV/$c$ and $Q_x<3$~MeV/$c$. These
``allowed'' combinations are obtained from a dedicated simulation
using the precise geometry of the setup.  If a relevant combination is
found, the T4 processor generates a positive decision signal which
starts the data transfer to the VME buffer memories.  Otherwise, the
Clear and Reset signals are applied to the DAQ and trigger systems.

The T4 decision time depends on the complexity of the event and is
around 3.5~$\mu$s on average. The rejection factor of T4 is around 5
with respect to the T1 rate and around 2.5 with respect to DNA/RNA.

The whole trigger system is fully computer controlled: no hardware
intervention is needed in order to modify the trigger configuration.
With all selection stages enabled the event rate at the typical
experimental conditions is around 700 per spill, that is well below
the limits of the DAQ rate capability.

\section{Data acquisition system}

The architecture of hardware and software parts of the data
acquisition system takes into account the time structure of the proton
beam. The machine super-cycle of the CERN PS accelerator has 15--20~s
duration. Within this period DIRAC receives protons in spills of
400--450~ms width, from 1 to 5 spills per super-cycle.  The interval
between consecutive spills can be as short as 1~s.

During the accelerator burst the data from all detectors are read out
into VME buffer memories (commercial and dedicated electronic modules)
without any software intervention.  The data transfer to VME processor
boards, event building, data transfer to the main host computer and
other relatively slow operations are performed during the pause
between bursts.  This provides the maximum operation rate of DAQ.  The
information comes from 2048 channels of microstrip gas chambers, 800
channels of scintillating fibre detector, 2016 channels of drift
chambers and 224 channels of other scintillation and Cherenkov
detectors. For every channel the time or amplitude information is
recorded or both.  In addition to a main readout mode, the readout of
scalers at the end of every spill is arranged via a CAMAC bus.  The
counting rates of all the detectors, trigger rates at different
trigger levels and for different sub-trigger modes are recorded
together with the value of the beam intensity provided by the PS
complex.

\subsection{DAQ hardware}

The data readout \cite{daqhard} is arranged with 12 readout branches:
4 branches for GEM/MSGC, 3 branches for DC and 5 FERA \cite{FERA}
branches for all other detectors. In FERA and DC branches the VME
modules CES\,HSM\,1870 and LeCroy\,1190 are used as buffer memories.
In GEM/MSGC branches the buffer memories are incorporated into
dedicated VME modules \cite{MSGCelectr}.  FERA branches include
different FERA compatible LeCroy modules: ADC\,4300B, multi-hit
TDC\,3377, universal logic modules 2366 configured in this application
like FERA registers and scalers. Peculiarities of FERA readout in
DIRAC, such as multi-gate and Fast Clear operation, are described in
\cite{FERAclear}.

The logic of readout is the following. The first level trigger T1
starts digitisation in ADC and TDC of FERA and DC branches (see
Fig.~\ref{structure} in Section ``Trigger system''). Readout is
inhibited unless a positive decision of the highest level trigger T4
is received.  If the T4 processor decides positively, the Inhibit
Readout status is released and the converted event data are
transferred to buffer memories.  If the decision of DNA/RNA or T4 is
negative, then a Fast Clear signal is generated \cite{FERAclear} which
discards the data in all FERA modules and DC registers.

In contrast to FERA and DC subsystems, the processing of GEM/MSGC data
is started by the next level DNA/RNA trigger. This reduces the dead
time introduced by the Clear process which in GEM/MSGC electronics
takes longer. Thus, the negative decisions of T4 only lead to clearing
of the GEM/MSGC data. If no Clear signal is received, the converted
data are transferred to buffer memories.

The readout of the whole event takes 45~$\mu$s and is defined by a
fixed acquisition time of GEM/MSGC which exceeds the readout time in
other branches.

\subsection{DAQ software}

The main part of the DAQ software \cite{daqsoft} is running on two VME
processor boards and on the main DAQ host. PowerPC-based VME processor
boards control the trigger and FERA readout electronics via two CAMAC
branch drivers, operate VME modules, read data from buffer memories
and transfer them to the main DAQ host. The main DAQ host performs
event building, records the built data and distributes them to other
computers for on-line monitoring and analysis.

The DAQ software is written in C programming language and is running
under UNIX-like operating systems: Lynx-OS (VME processors) and LINUX
(main DAQ host).

\begin{figure}[htb]
\begin{center}
  \includegraphics*[height=15cm,bb=95 130 525 661]{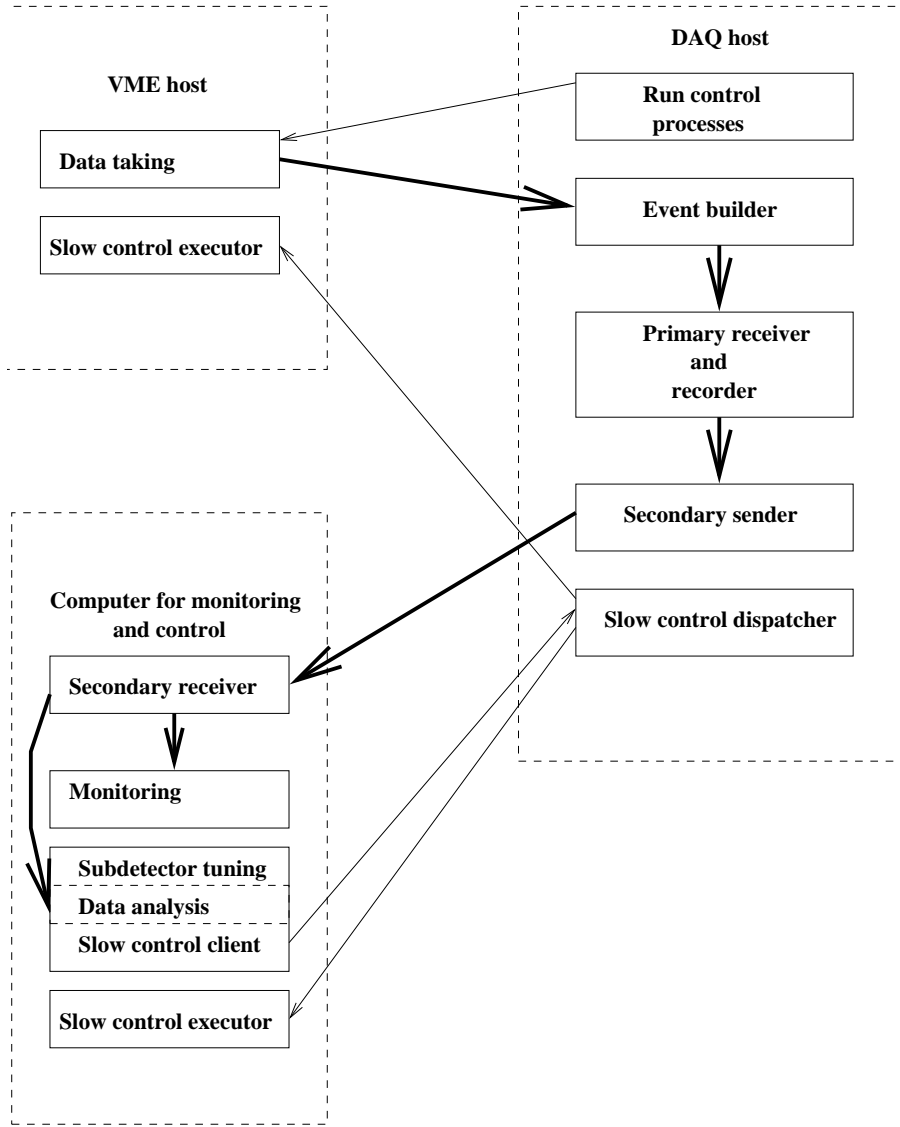}
\end{center}
\caption{Layout of the DAQ architecture.}
\label{soft}
\end{figure}

The schematic layout of the data acquisition processes is presented in
Fig.~\ref{soft}. The basic processes running on VME and main DAQ hosts
and on one of the monitoring computers are shown. Only one VME
processor is shown for simplicity.

The DAQ software can be divided into more or less independent groups.
The first one is a set of programs for data readout (runs on VME
boards) and event building (runs on the main DAQ host). During the
event building the data are checked for consistency using the headers
and the serial numbers of sub-events provided by the detector readout
electronics. The serial numbers allow to check that all buffer
memories contain the same number of sub-events and that different
sub-events correspond to each other.

The second group consists of programs for data distribution over the
network for on-line monitoring. The group includes a primary data
receiver and secondary data senders/receivers. The primary receiver is
the only program which receives data directly from the event builder
and is critical for the data acquisition: other processes of this
group may either be or not be running and may be started or stopped at
any time.

The third group of programs allows to control the run status via a
graphical user interface: to set the run parameters, to select the
trigger type, to change the run status (start, stop, suspend etc.).
This group includes also a Run Display program which delivers
information about the current run conditions.

The fourth group is a set of slow-control processes providing a
uniform access to hardware from any host participating in DAQ. The
tasks of these programs are, for example, high voltage control, VME
and CAMAC module initialisation and so on.

The whole DAQ system is able to accept data from five consecutive
bursts in one accelerator super-cycle with up to 2~MBytes per burst
(this corresponds to ~2000 triggers per burst) and to distribute
events to all hosts participating in on-line data processing. The
limitation comes from the capacity of the VME buffer memories.

\subsection{On-line monitoring}

An on-line monitoring program receives the data distributed by the
main DAQ process. The program fulfils accumulation of several hundred
histograms, including all raw signal spectra from any detector as well
as spectra obtained after a fast preliminary data handling. The
software is written in the frame of ROOT software package.

The program delivers the following information:
\begin{itemize}
\item[---] Time and amplitude distributions for every counter.
\item[---] Hit and multiplicity distributions.
\item[---] Correlation plots for hits in different detectors.
\item[---] Beam profile on the experimental target, beam spill time structure.
\item[---] Some specific data for selected detectors (cluster size
  for GEM/MSGC etc.).
\item[---] Event rates for different trigger modes running in parallel
  and their timing.
\item[---] Information about the data taking process as a whole:
  burst data volume, event volume, time of event in burst etc.
\end{itemize}
Proper setting of the corresponding program parameter allows selection
of the trigger type for which all the above operations should be
fulfilled. For example, the histograms can be obtained separately for
$e^+e^-$, $\pi^+ \pi^-$, $\Lambda$ and any other trigger or for all
accepted triggers.

Apart from the common program, there are dedicated monitoring programs
developed for more detailed control of individual detectors.

\section{Setup performances}

The experimental setup is designed to provide high efficiency for
detection of $\pi^+\pi^-$ pairs with small opening angle in the
laboratory frame ($\theta <$ 3~mrad) and small relative momentum in
the center of mass frame ($Q <$ 3~ MeV/$c$). Detector resolution and
multiple Coulomb scattering in the target and setup elements affect
the measurement accuracy.

Detector resolution leads to a linear increase of the relative error
$\sigma_p/p$ with increasing momentum value, whereas multiple
scattering contributes a constant term to the momentum uncertainty.
With an experimental space resolution of 90~$\mu$m in the DC system
and 50~$\mu$m in the GEM/MSGC detectors, the uncertainty on the
coordinate and momentum measurements are mainly determined by the
description of the multiple scattering errors.

Both momentum and the opening angle measurement accuracies affect the
precision on the measurement of the relative momentum $Q$ of the pion
pair.  The momentum accuracy $\sigma_p/p$ is 0.3$\%$, at $p_{\pi} =
p_{A_{2\pi}}/2$ = 2~ GeV/$c$, almost independent on momentum.
 
The kinematics of $\pi^+\pi^-$ pairs originating from the ionisation
of $A_{2\pi}$ atoms in the target requires ability on two-track
separation in the upstream detector region better than 1.5~mm in most
of the cases. The experimental double track resolution is limited by
the granularity and signal clustering of the upstream detectors and
with current algorithms is about 0.4~mm. Thus, the accuracy on the
measurement of the longitudinal ($Q_L$) and transversal components
($Q_x$ and $Q_y$) of $Q$ is 0.6~MeV/$c$ ($Q_L$) and 0.4~MeV/$c$ ($Q_x$
and $Q_y$).  These errors arise from the estimated contribution of
multiple scattering in the detectors and setup elements, according to
our present knowledge of the experimental data.  In fact, the effect
of multiple scattering in the foil target increases the error in $Q_x$
and $Q_y$ to up 1~MeV/$c$.

\begin{figure}[htb]
\begin{center}
  \includegraphics*[width=12cm,height=8cm]{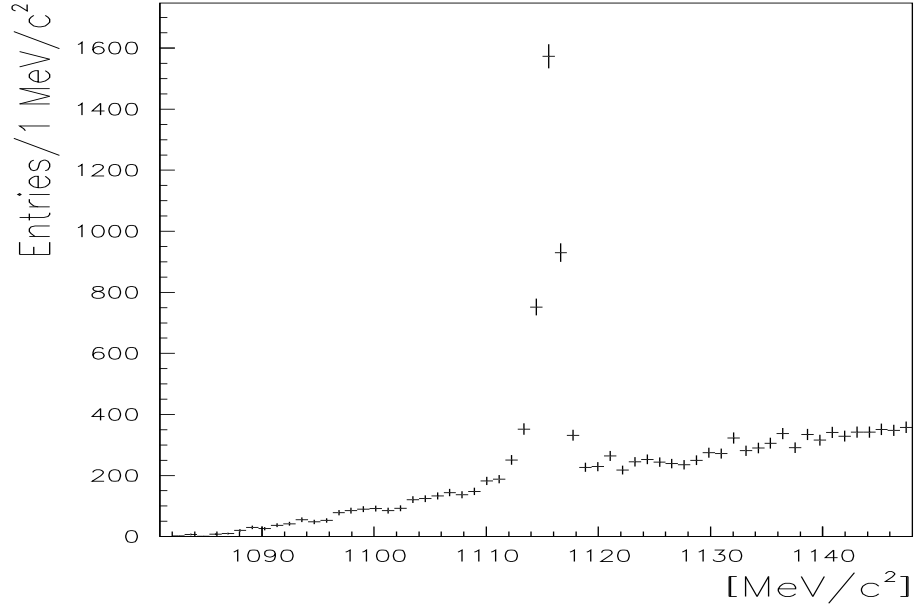}
\end{center}
\caption{Invariant mass distribution of $p \pi^-$ pairs. Events were selected
  from a sample of $\Lambda$ calibration data.}
\label{lambda}
\end{figure}

The calibration of the energy scale and the experimental determination
of the setup resolution are performed by monitoring the position and
the width of the $\Lambda$ detected by the apparatus. The
reconstructed $p \pi^-$ invariant mass obtained from calibration data
is shown in Fig.~\ref{lambda}.  The $\Lambda$ is clearly seen above a
small background. A gaussian plus a polynomial fit to the data gives
$M_{\Lambda}$ = 1115.67~MeV/$c^2$ and $\sigma_{\Lambda} =
0.43$~MeV/$c^2$. Calibration using $\Lambda \rightarrow p \pi^-$
decays is used in addition to control the precision of the setup
alignment. Any misalignment of the tracking system in one arm relative
to the other arm would result in asymmetrical errors on the
reconstructed momenta.

\begin{figure}[htb]
\begin{center}
  \includegraphics*[width=\textwidth,height=6cm,bb=12 24 522
  230]{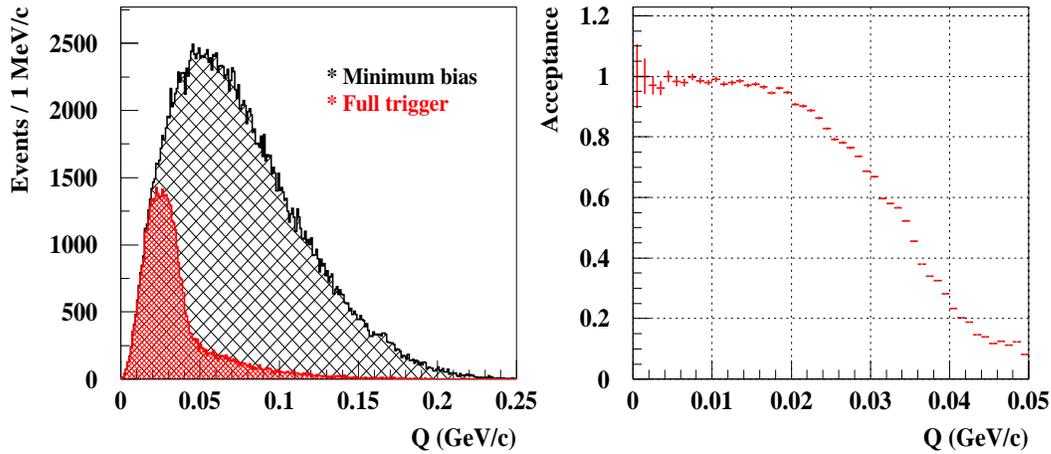}
\end{center}
\caption{Left: Distribution of $Q$ for accepted $\pi^+\pi^-$ pairs
  after the full DIRAC trigger system, and for minimum bias pairs.
  Right: trigger acceptance, determined as the ratio between the
  previous two distributions.}
\label{trigacc}
\end{figure}

\begin{figure}[htb]
\begin{center}
  \includegraphics*[width=\textwidth,height=0.4\textwidth]{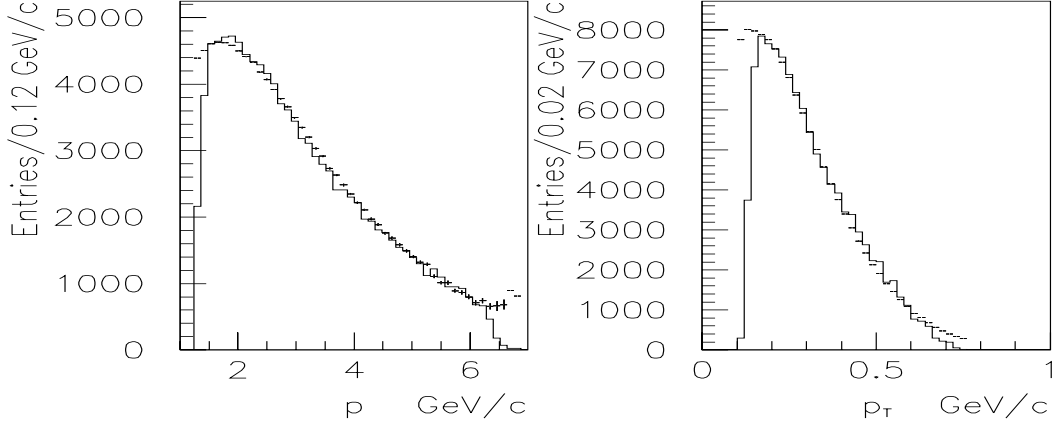}
\end{center}
\caption{Experimental $p$ and $p_T$ distributions of $\pi^-$ corrected
  for apparatus acceptance (histogram) with superimposed the results
  of the parameterisation \cite{Badh}\cite{lana}.}
\label{incl_spectra}
\end{figure}

The accuracy of the time measurement is obtained from the analysis of
the distribution of the $e^+ e^-$ time difference at the vertical
hodoscopes as explained in that section (see Fig.~\ref{e+e-}).

The performance of the trigger system as a whole in selecting low-Q
events is illustrated in Fig \ref{trigacc}, where the magnitude of
relative momentum of pion pairs $Q$ is shown (in their center-of-mass
frame), after DNA/RNA and T4 trigger selection (see trigger section).
Trigger efficiency as function of $Q$ is flat in the low-Q region, as
illustrated in Fig.~\ref{trigacc}.  This is considered an important
figure of merit of the spectrometer, for a precision study of the
$\pi^+ \pi^-$ Coulomb interaction.
  
Pions from ionisation of $A_{2\pi}$ entering the apparatus have
momenta below 4~GeV/$c$.  The apparatus momentum acceptance for
time-correlated pairs is flat for pions with momenta between
1.6~GeV/$c$ and 3~GeV/$c$, and it decreases for higher momenta.

For the sake of completeness we also show the $p$ and $p_T$
distributions for a single $\pi^-$ in Fig.~\ref{incl_spectra}, with
superimposed a parameterisation of the inclusive yield based on the
analytic representation \cite{Badh}, adapted to DIRAC center-of-mass
energy ($\sqrt s$=6.84~GeV) \cite{lana}.

\section*{Acknowledgements}

We would like to acknowledge the financial support received by the
institutions in the DIRAC collaboration for the construction of the
spectrometer, in particular by the following funding agencies: the
Swiss National Science Foundation, the Ministerio de Ciencia y
Tecnologia (Spain), under projects AEN96-1671 and AEN99-0488, the
PGIDT of Xunta de Galicia (Spain), the Istituto Nazionale di Fisica
Nucleare (Italy), the Ministery of Industry, Science and Technologies
of the Russian Federation and the Russian Foundation for Basic
Research (Russia), under project 01-02-17756, the IN2P3 (France),the
Greek General Secretariat of Research and Technology (Greece), the
University of Ioannina Research Committee (Greece), the Grant Agency
of the Czech Republic, grant No. 202/01/0779, and the Japan Society
for the Promotion of Science (JSPS), Grant-in-Aid for Scientific
Research No.  07454056, 08044098, 09640376, 11440082, 11694099,
12440069 and 14340079.

We wish to thank the CERN Directorate for the continuos encouragement
and support to DIRAC.  We also thank the personell of the CERN PS
Division for their essential contribution to the experiment. We warmly
acknowledge the help of J.~Bosser, A.~Braem, M.~Bragadireanu,
N.~Chritin, L.~Danloy, M.~Doser, B.~Dulach, L.~Durieu, O.~Ferrando,
W.~Flegel, A.~Froton, J.~Ch.~Gayde, P.~A.~Giudicci, M.~Hauschild,
J.~Y.~Hemery, M.G.~Iovanozzi, G.~Martini, G.~Molinari, J.M.~Nonglaton,
V.~Prieto, J.P.~Riunaud, T.~Ruf, D.~Simon, R.~Steerenberg,
Ch.~Steinbach, J.W.N.~Tuyn, M.~Zahnd and M.~Zanolli.

We are also grateful to the Directorate of Joint Institute for Nuclear
Research, for its support to the experiment.

Finally would like to thank our secretary C.~Moine for the care she
has put in solving our every-day problems.

\end{document}